\documentclass{aa}  

\usepackage{natbib}
\usepackage{graphicx}
\usepackage{txfonts}
\usepackage{hyperref}

\usepackage{xcolor}

\newcommand{\planck}{\textsl{Planck}}

\newcommand{\rosat}{\textsl{ROSAT}}

\newcommand{\suzaku}{\textsl{SUZAKU}}
\newcommand{\erosita}{\textsl{eROSITA}}

\newcommand{\beq}{\begin{equation}}
\newcommand{\eeq}{\end{equation}}
\newcommand{\beqa}{\begin{eqnarray}}
\newcommand{\eeqa}{\end{eqnarray}}
\newcommand{\mpc}{$h^{-1} \mathrm{Mpc}$}

\newcommand{\ie}{i.e.\xspace}

\def\msun{{\rm M}_{\odot}}
\def\der{{\rm d}}

\usepackage{comment}

\usepackage{amstext}

\begin{document} 

    \title{Density and temperature of cosmic-web filaments \\ on scales of tens of megaparsecs}
  \author{H. Tanimura \and N. Aghanim \and V. Bonjean \and N. Malavasi \and M. Douspis}

  \institute{
  Universit\'{e} Paris-Saclay, CNRS, Institut d'astrophysique spatiale, 91405, Orsay, France \\
    \email{hideki.tanimura@ias.u-psud.fr}
             }
             
  \date{}
  
  \abstract 
    {We studied physical properties of matter in 24,544 filaments ranging from 30 to 100 Mpc in length, identified in the Sloan Digital Sky Survey (SDSS). We stacked the Comptonization $y$ map produced by the \planck\ Collaboration around the filaments, excluding the resolved galaxy groups and clusters above a mass of $\sim3 \times 10^{13} \, \msun$. We detected the thermal Sunyaev-Zel'dovich signal for the first time at a significance of 4.4$\sigma$ in filamentary structures on such a large scale. We also stacked the \planck\ cosmic microwave background (CMB) lensing convergence map in the same manner and detected the lensing signal at a significance of 8.1$\sigma$. To estimate physical properties of the matter, we considered an isothermal cylindrical filament model with a density distribution following a $\beta$-model ($\beta$=2/3). Assuming that the gas distribution follows the dark matter distribution, we estimate that the central gas and matter overdensity $\delta$ and gas temperature $T_{\rm e}$ are $\delta = 19.0^{+27.3}_{-12.1}$ and $T_{\rm e} =  1.4^{+0.4}_{-0.4} \times 10^6$ K, which results in a measured baryon fraction of $0.080^{+0.116}_{-0.051} \times \Omega_{\rm b}$.}

\maketitle


\section{Introduction}
\label{sec:intro}

The large-scale structure of the Universe is organized in a web-like structure called the cosmic web \citep{Bond1996}. The formation of the cosmic-web structure is described in the standard cosmological model of structure formation (e.g., \citealt{Zeldovich1982}). The cosmic-web structure is governed by the gravitational collapse of matter in the expanding Universe, therefore it can be used to test cosmological models. 

Numerical simulations illustrate that the cosmic web consists of nodes, filaments, sheets, and voids. For example, nodes form in dense regions, which are interconnected by filaments and sheets, and most regions are occupied by voids with extremely low density. Galaxy clusters are prominent structures that are easily identified in the large-scale structure, but it is predicted that the largest fraction of matter (around 50\% in mass) resides in filamentary structures \citep{cen2006,Aragon2010, Cautun2014}. 

Observationally, the filamentary structures are identified by the distribution of galaxies in large surveys such as the 2dF galaxy redshift survey (2dFGRS) \citep{Colless2003}, two micron all sky survey (2MASS) \citep{Skrutskie2006}, sloan digital sky survey (SDSS) \citep{Gunn2006}, and VIMOS public extragalactic redshift surve (VIPERS) \citep{Scodeggio2018}. However, hydrodynamical simulations predict that galaxies only comprise less than 10\% of baryonic matter and that the majority of baryons exist in filaments in the form of hot plasma: a temperature range of $10^5$--$10^7$ K and densities of about 10--100 times the average cosmic value, which is referred to as the warm- hot intergalactic medium (WHIM) \citep{Cen1999, cen2006, Martizzi2018}. Extensive search for the WHIM has been performed in far-UV, X-ray, and with the thermal Sunyaev-Zel'dovich (tSZ) effect  \citep{Zeldovich1969, Sunyaev1970, Sunyaev1972}. However, the measurement is difficult because the signal is relatively weak and the morphology of the source is complex. 

Several detections of gas in filaments have been reported based on X-ray and tSZ observations; for example, filaments between nearby pairs of galaxy clusters have been studied. Extensive studies have been performed for the filament between the pair of A399 and A401 \citep{Fujita1996, Fujita2008, Planck2013IR-VIII, Bonjean2018} and for the filament connecting the massive Abell clusters A222 and A223 \citep{Dietrich2005, Werner2008}, as well as for the filament between A3391 and A3395 \citep{Tittley2001, Planck2013IR-VIII, Alvarez2018, Sugawara2017}.  Individually, filaments extending over the virial radii of Abell 133 \citep{Vikhlinin2013, Connor2018, Connor2019}, Abell 2744 \citep{Eckert2015}, and A1750 \citep{Bulbul2016} were thoroughly studied, and hot dense gas was found that is connected to the large-scale structure.  However, the detections are limited to relatively short ($\sim$ 10 \mpc\ ) and highly dense detections ($\delta > 150$), especially for individual filaments.

Statistical detections of gas or matter in filaments have also been reported. The lensing signal was detected by stacking filaments between pairs of luminous red galaxies (LRGs) \citep{Clampitt2016, Epps2017, Xia2019} and pairs of the baryon oscillation spectroscopic survey (BOSS) constant-mass (CMASS) galaxies \citep{Kondo2019}, as well as by cross-correlating the filaments in \cite{Chen2016} with lensing of the cosmic microwave background (CMB) \citep{He2018}, or by studying the azimuthal distribution of galaxies around galaxy clusters \citep{Gouin2019}. In addition, the tSZ signal was detected by stacking filaments between pairs of LRGs \citep{Tanimura2019b} and pairs of CMASS galaxies \citep{deGraaff2019}. These measurements estimate a gas or matter density of $\delta = 3 \sim 5$, which is much lower than the values estimated for the detected individual objects. Despite these various detections, most of the results are still limited to short filaments of about 10 \mpc\ .

To investigate longer filaments with more complex morphologies, it is crucial to identify their precise locations and shapes. Several algorithms have been developed for this purpose  (e.g., \citealt{Aragon2007, Aragon2010, Aragon2010b, Forero2009, Sousbie2011, Falck2012, Cautun2013, Stoica2005, Bonnaire2019}). Some of these methods were applied to actual data, such as the Bisous method, which was applied to SDSS-DR8 galaxies in \cite{Tempel2014} to construct a filament catalog, or the subspace-constrained mean shift (SCMS) algorithm, which was applied by \cite{Chen2016} to SDSS-DR12 galaxies to detect locations of filaments. 

The discrete persistent structure extractor (DisPerSE) algorithm \citep{Sousbie2011} has been applied to SDSS-DR12 galaxies in \cite{Malavasi2020arXiv} to construct a filament catalog. \cite{Bonjean2019} confirmed the galaxy distribution around the filaments and studied the galaxy properties such as star formation activity around the filaments. In the present study, we estimate gas and matter properties such as density and temperature in the filaments. 

This paper is organized as follows: Section \ref{sec:data} summarizes the data sets we used in our analyses. Section \ref{sec:ana} explains our stacking method for measuring the average structure around filaments using the tSZ and CMB lensing data. Possible systematic errors in our measurements are discussed in Section \ref{sec:systematics}. Section \ref{sec:model} explains the model with which we interpret our measurements, which is followed by our results in Section \ref{sec:result}.   We end the paper with discussions and conclusions in Section \ref{sec:discussion} and Section \ref{sec:conclusion}. Throughout this work, we adopt the $\Lambda$CDM cosmology from \cite{planck2016-xiii} with $\Omega_{\rm m} = 0.3075$, $\Omega_{\Lambda} = 0.6910$, $\Omega_{\rm b} = 0.0486$, and $H_0 = 67.74$ km s$^{-1}$ Mpc$^{-1}$. All masses are quoted in solar mass, and $M_{\Delta}$ is mass enclosed within a sphere of radius $R_{\Delta}$ such that the enclosed density is $\Delta$ times the {\it critical} density at redshift $z$. Uncertainties are given at the 1$\sigma$ confidence level.


\section{Data}
\label{sec:data}

\subsection{Planck y maps}
\label{subsec:ymap}

\planck\ all-sky $y$ maps are provided in the \planck\ 2015 data release\footnote{https://pla.esac.esa.int} in HEALpix\footnote{http://healpix.sourceforge.net/} format \citep{gorski2005} with a pixel resolution of $N_{\rm side}$ = 2048 ($\sim 1.7$ arcmin). The $y$ maps are constructed with two methods: the modified internal linear combination algorithm (MILCA) \citep{hurier2013} and needlet independent linear combination (NILC) \citep{remazeilles2013}, by combining multiband \planck\ frequency maps \citep{Planck2016-XXII} so that the spectral response of the Compton $y$ parameter is unity. We have found that our results are consistent using these two $y$ maps, and we show the results with the MILCA $y$ map in this paper. 

\begin{figure}
    \centering
    \includegraphics[width=\linewidth]{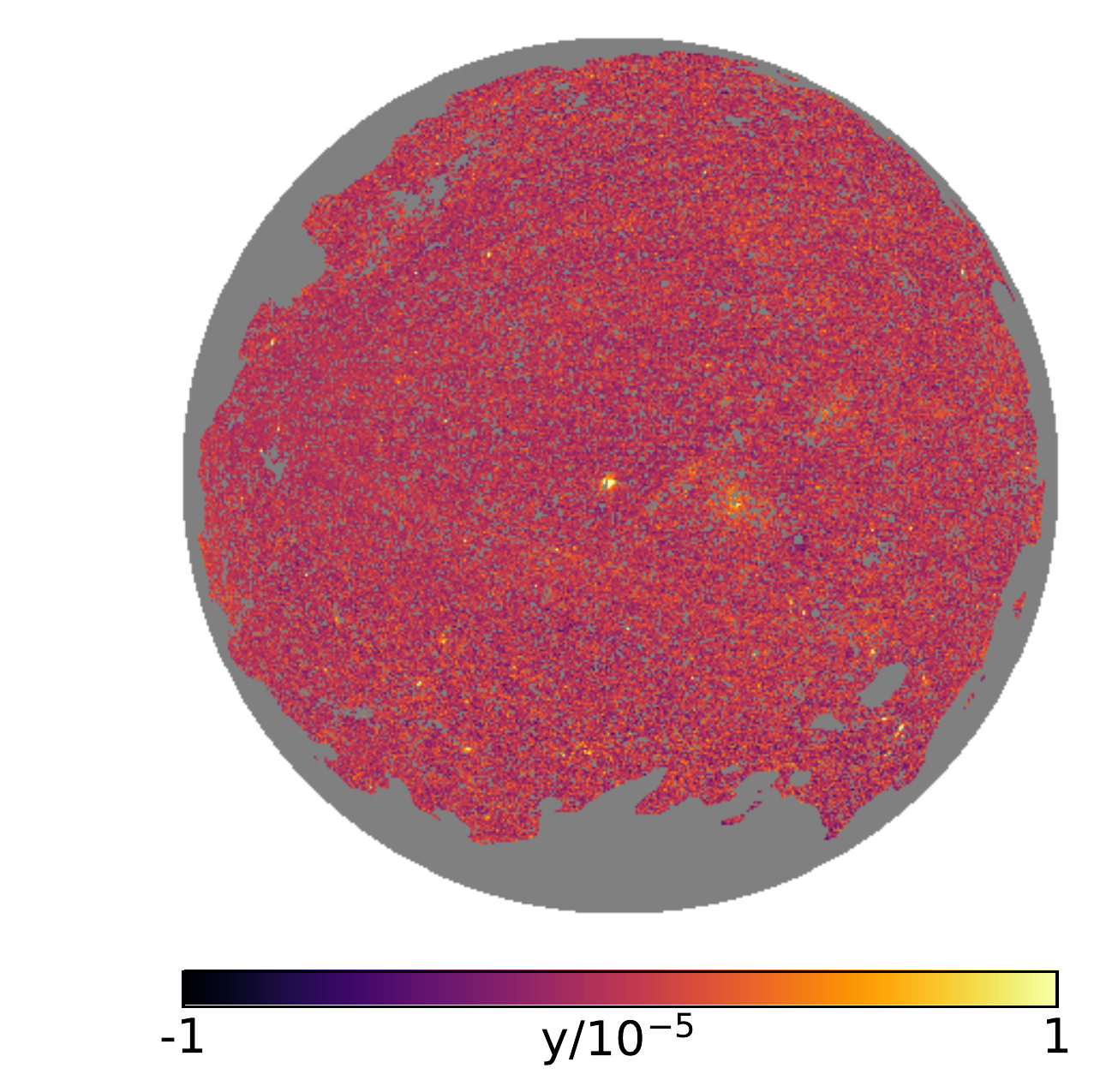}
    \includegraphics[width=\linewidth]{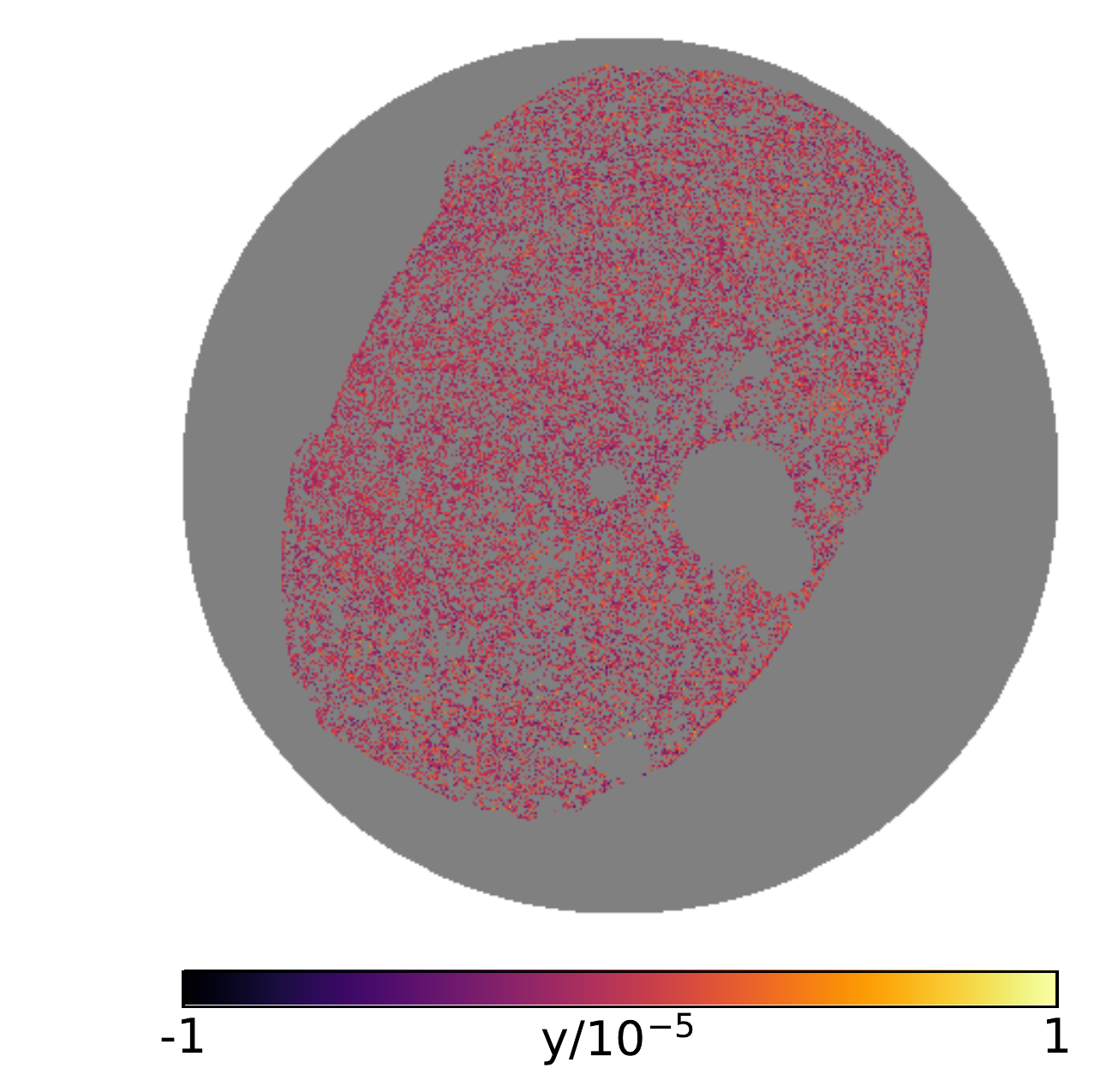}
    \caption{{\it Upper}: \planck\ $y$ map with the 40\% Galactic mask and the point-source mask viewed from the north Galactic pole. {\it Lower:} \planck\ $y$ map after masking galaxy clusters. The galaxy groups and clusters detected by the \planck\ tSZ, \rosat\ X-ray, and SDSS optical surveys described in Sect. \ref{subsec:cluster-catalog} are all masked by three times the radius ($3 \times R_{500}$). In addition, critical points (maxima and bifurcations) from DisPerSE, with overdensities exceeding 5, are masked within a radius of 10 arcmin. Regions outside the SDSS DR12 survey are also excluded. }
    \label{ymap}
    \end{figure}

The 2015 \planck\ data release also provided sky masks that are suitable for the analysis of the $y$ maps, masking point sources that cover about 10\% of the sky, and regions around the Galactic plane that exclude 40, 50, 60, and 70\% of the sky. We combined the point-source mask and the 40\% Galactic mask, which excludes about 50\% of the sky ({\it upper panel} in Fig.~\ref{ymap}).

\subsection{Planck CMB lensing map}
\label{subsec:kmap}
\begin{figure}
    \centering
    \includegraphics[width=\linewidth]{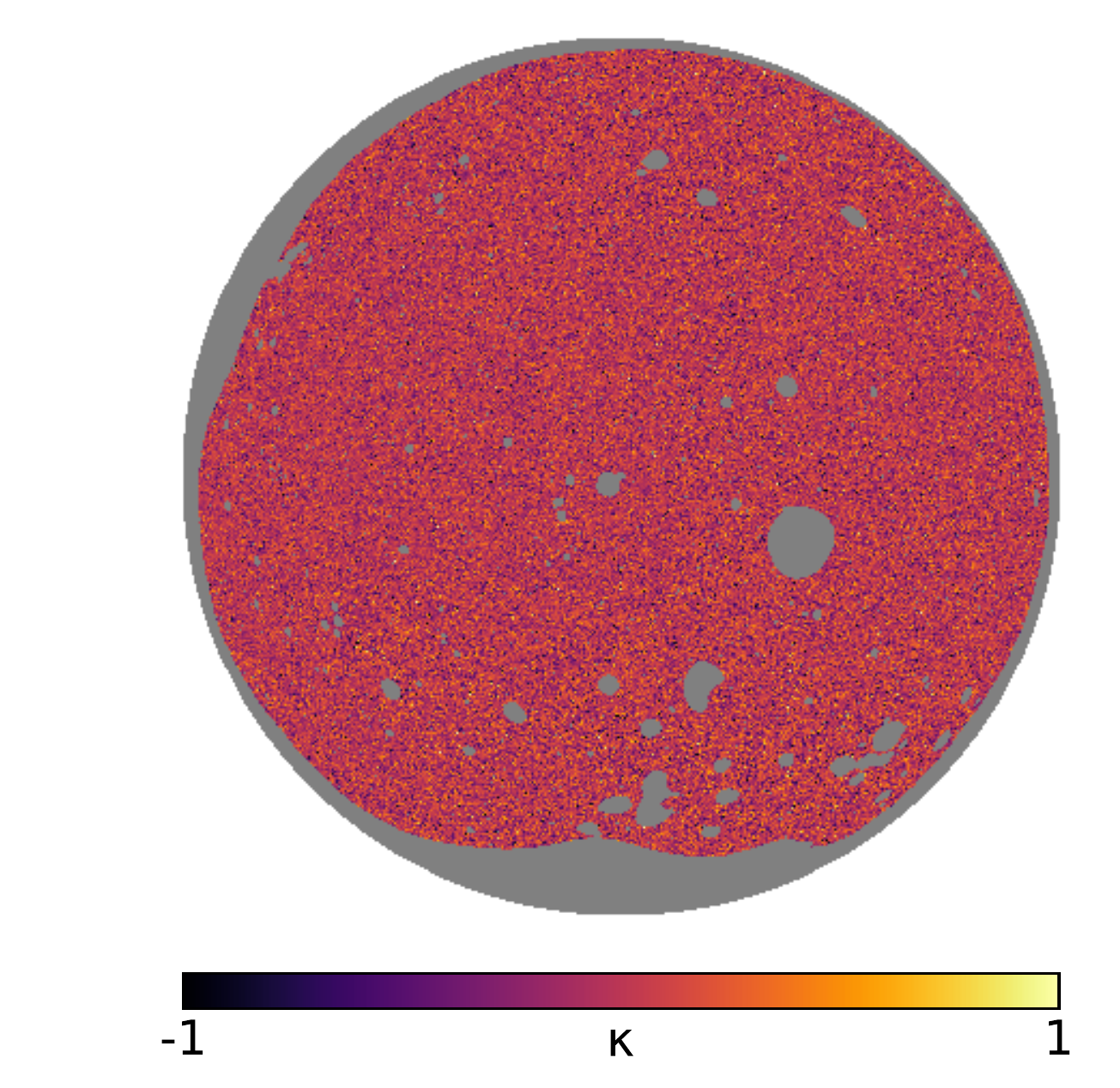}
    \includegraphics[width=\linewidth]{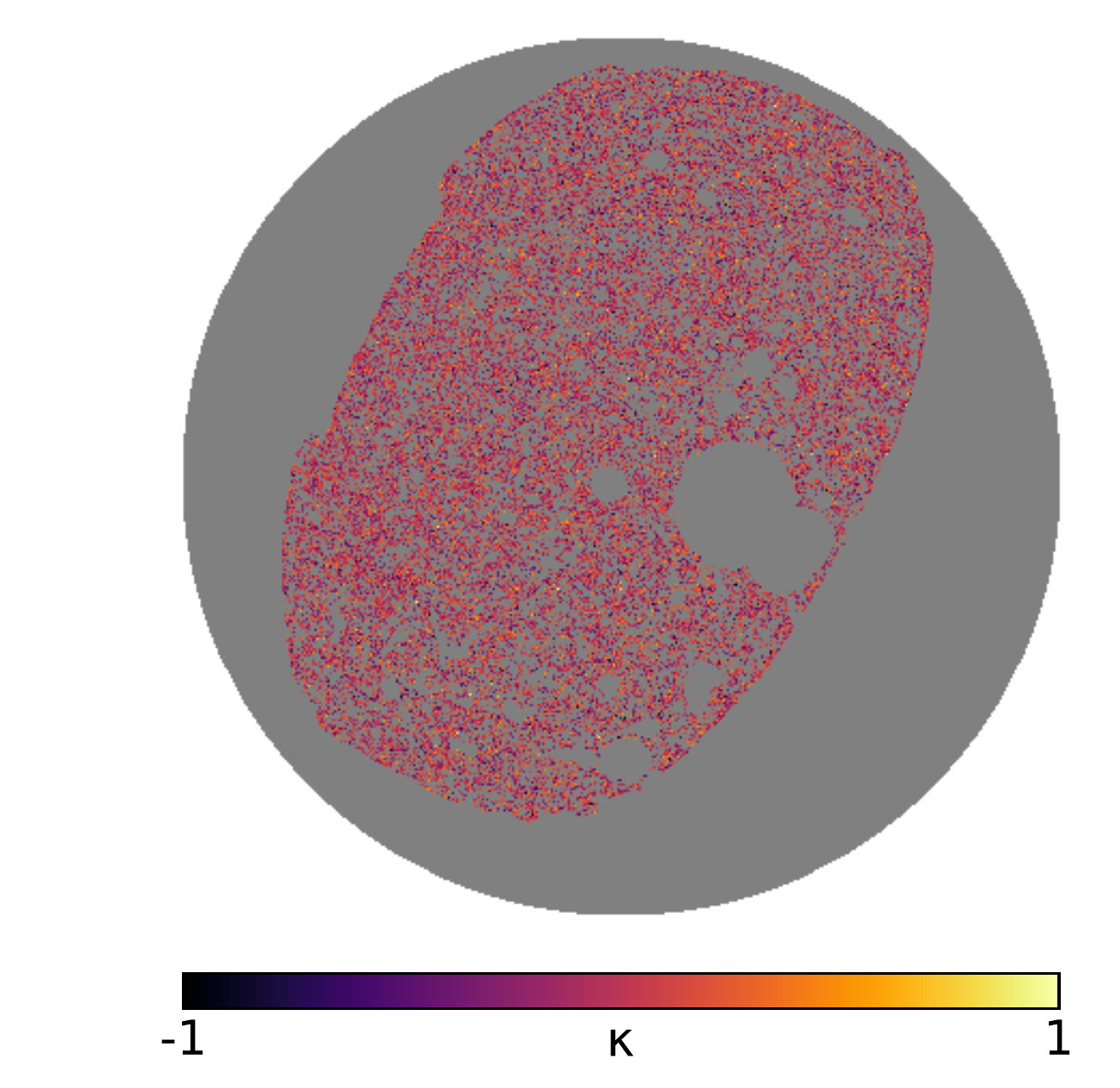}    
    \caption{{\it Upper}: \planck\ CMB lensing map with the mask provided by the \planck\ team viewed from the north Galactic pole. {\it Lower:} \planck\ CMB lensing map after masking galaxy clusters, critical points from DisPerSE, and regions outside the SDSS DR12 survey (same mask as in the lower panel in Fig.~\ref{ymap}). }
    \label{kmap}
    \end{figure}

The final \planck\ 2018 data products provide the \planck\ all-sky CMB lensing map \citep{planck2018-viii}. The convergence($\kappa$) reconstruction has been performed up to $\ell_{max}$ = 4096 from the CMB map based on the spectral matching independent component analysis (SMICA). Three baseline lensing maps are provided from temperature-only (TT), polarization-only (PP), and minimum-variance estimates from the temperature and polarization (MV). The \planck\ team also provides the sky mask that is suitable for the analysis of the $\kappa$ map, masking point sources and regions around the Galactic plane as well as the \planck\ 2015 SZ clusters, leaving about 67\% of the sky unmasked. In our study, we use the CMB lensing map of the MV estimate, with which a variety of tests have been performed in \cite{planck2018-viii}.  The map shows a strong increase in noise at small scales, therefore we used the conservative multipole range ($\ell < 400$), as recommended by the \planck\ team in \cite{planck2018-viii}. We filtered the higher multipole range ($\ell > 400$) with an exponential function ({\it upper panel} in Fig.~\ref{kmap}).

\subsection{Filament catalog}
\label{subsec:filament-catalog}
We used a sample of filaments selected from the filament  catalog  constructed  by  \cite{Malavasi2020arXiv} with the DisPerSE algorithm, to which we refer for details on the filament catalog. The DisPerSE method computes the gradient of  the  density  field  and  identifies  critical  points where the gradient is zero. These critical points can then be classified as maxima of the density field (associated with groups or clusters of galaxies), minima, and saddles (local density minima bounded to structures, such as filaments). DisPerSE then defines filaments as field lines of constant gradient that connect critical points (maxima and saddles) and also computes the lengths between the critical points. Other critical points, called bifurcation points, are defined as the positions where filaments intersect. \cite{Malavasi2020arXiv} applied DisPerSE to the LOWZ-CMASS spectroscopic galaxies of \cite{Reid2016}, which are stated to be 99\% (97\%) complete for CMASS (LOWZ). The authors showed (Fig 11 of \cite{Malavasi2020arXiv}) that the spatial distribution of filaments is uniform throughout the sky. In our analysis, the galaxy density field used in the DisPerSE algorithm was calculated without any smoothing (zero-smoothing case in \cite{Malavasi2020arXiv}), and the persistence is $3\sigma$, where the persistence is the robustness ratio derived from the density ratio of critical points at both ends of filaments. For a given smoothing, the higher the persistence, the more reliable the detected filaments. As shown in \cite{Sousbie2011}, the fraction of false filaments detected with DisPerSE is about 5\% for a $2\sigma$ persistence threshold and about 0.006\% for a $4\sigma$ persistence. The filament catalog we used is constructed with a $3\sigma$ persistence threshold, implying that more than 95\% of the filaments are real. The DisPerSE algorithm outputs the positions of filaments as a collection of discrete segments. It also provides the positions and densities of the critical points that are identified in the galaxy density field.

\begin{figure}
    \centering
    \includegraphics[width=\linewidth]{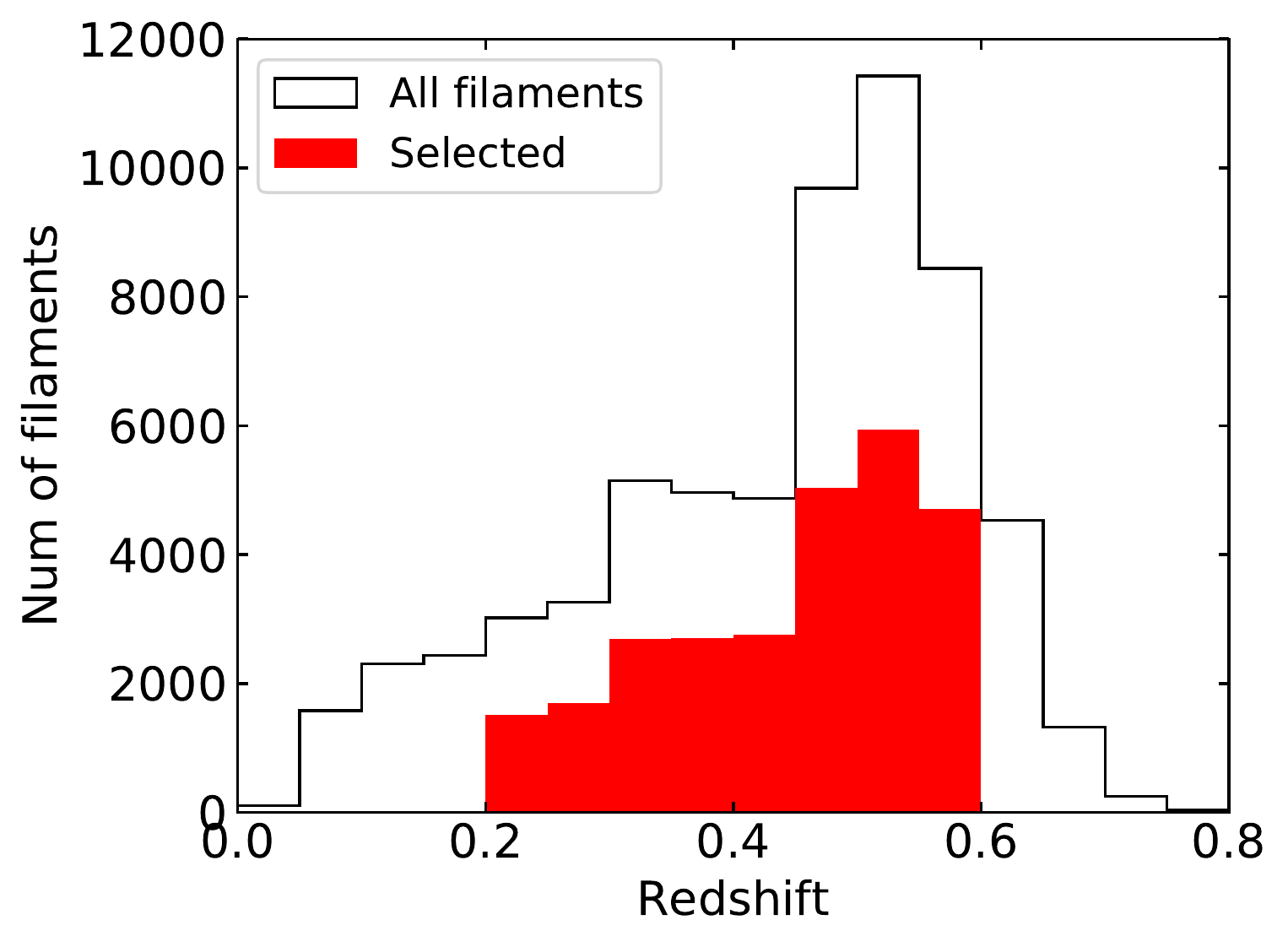}
    \includegraphics[width=\linewidth]{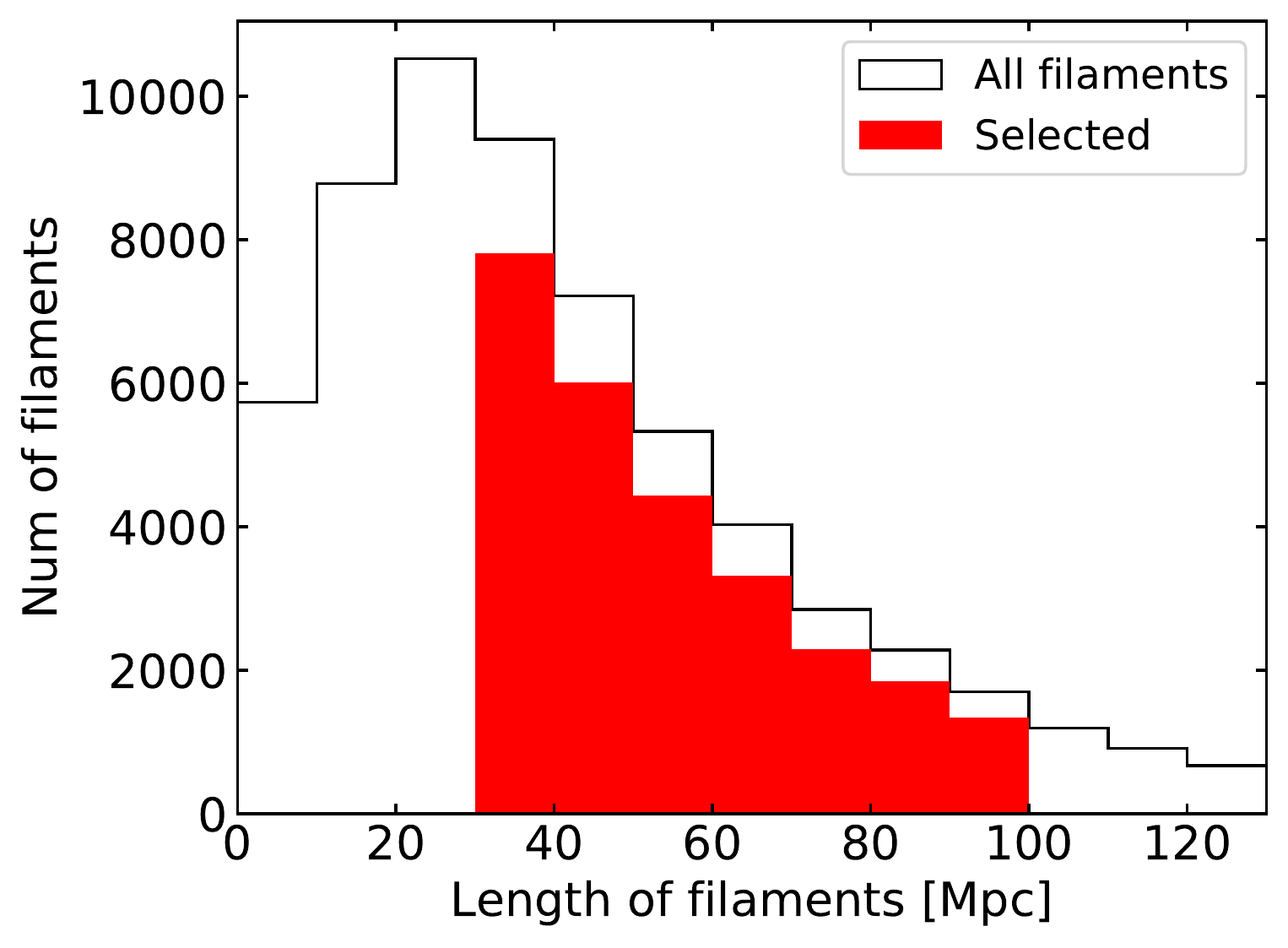}
    \caption{{\it Upper}: Redshift distribution of filaments. {\it Lower}: Length distribution of filaments. The {\it black line} shows the distribution of all the 63,391 filaments in the filament catalog, and {\it red line} is the distribution of 24,544 filaments used in this paper. }
    \label{fil-z-len}
    \end{figure}

Figure.~\ref{fil-z-len} shows the length and redshift distributions of the filaments. The {\it black line} is the distribution of all the 63,391 filaments in the parent catalog from \cite{Malavasi2020arXiv}. The {\it red line} shows the selected sample of filaments used in our analysis: 24,544 filaments with length 30 to 100 Mpc at $0.2<z<0.6,$ where the galaxy density is the most uniform. 

We selected filaments with lengths of 30--100 Mpc for the purpose of probing cosmic filaments that connect nodes in the large-scale structure. For this purpose, we removed short filaments that may link pairs of clusters or reside inside local structures such as filaments in superclusters. To determine a characteristic scale between nodes, we referred to the two-point correlation function of galaxy groups and clusters. \cite{yang2005} showed that the correlation function of galaxy groups with an average mass of $\sim1.2 \times 10^{13} \, h^{-1} \msun$ has a characteristic distance between 4--15 \mpc\ ($\sim$6--22 Mpc). However, according to \cite{Cautun2014}, low-mass halos with a mass $M_{200} < 5 \times 10^{13} \, h^{-1} \msun$ typically reside inside filaments. This implies that the distance between nodes should be larger than the distance in \cite{yang2005}. \cite{basilakos2004} also showed that the correlation function of rich Abell clusters has a distance of $\sim$21 \mpc\ ($\sim$30 Mpc). Therefore we removed filaments shorter than 30 Mpc from our study to separate cosmic filaments from short filaments that reside in local structures. In our analysis, we also removed filaments that are longer than 100 Mpc because they may be unreliable. 

\subsection{Catalog of galaxy groups and clusters}
\label{subsec:cluster-catalog}
In order to remove the contribution from foreground or background galaxy groups and clusters, the objects listed in \cite{Tanimura2019a} were masked. These include 1653 tSZ clusters from the \planck\ observations \citep{Planck2016-XXVII} and 1743 MCXC X-ray clusters from the \rosat\ X-ray survey \citep{Piffaretti2011} as well as 26,111 redMaPPer clusters \citep{Rykoff2014}, 158103 WHL clusters \citep{Wen2012, Wen2015}, and 46,479 AMF clusters \citep{Banerjee2018} detected from SDSS galaxies. Mass and redshift distributions of the groups and clusters are shown in \cite{Tanimura2019a}, and the mass distribution in the union catalog reaches down to $\sim3 \times 10^{13} \, \msun$. 


\section{Data analysis}
\label{sec:ana}

In this section we describe our procedure of stacking the \planck\ $y$ and $\kappa$ map at the filament positions. We also estimate the signal associated with filaments together with the signal uncertainty. 

\subsection{Stacking maps at the filament positions}
\label{subsec:ystacking}
\begin{figure}
    \centering
    \includegraphics[width=\linewidth]{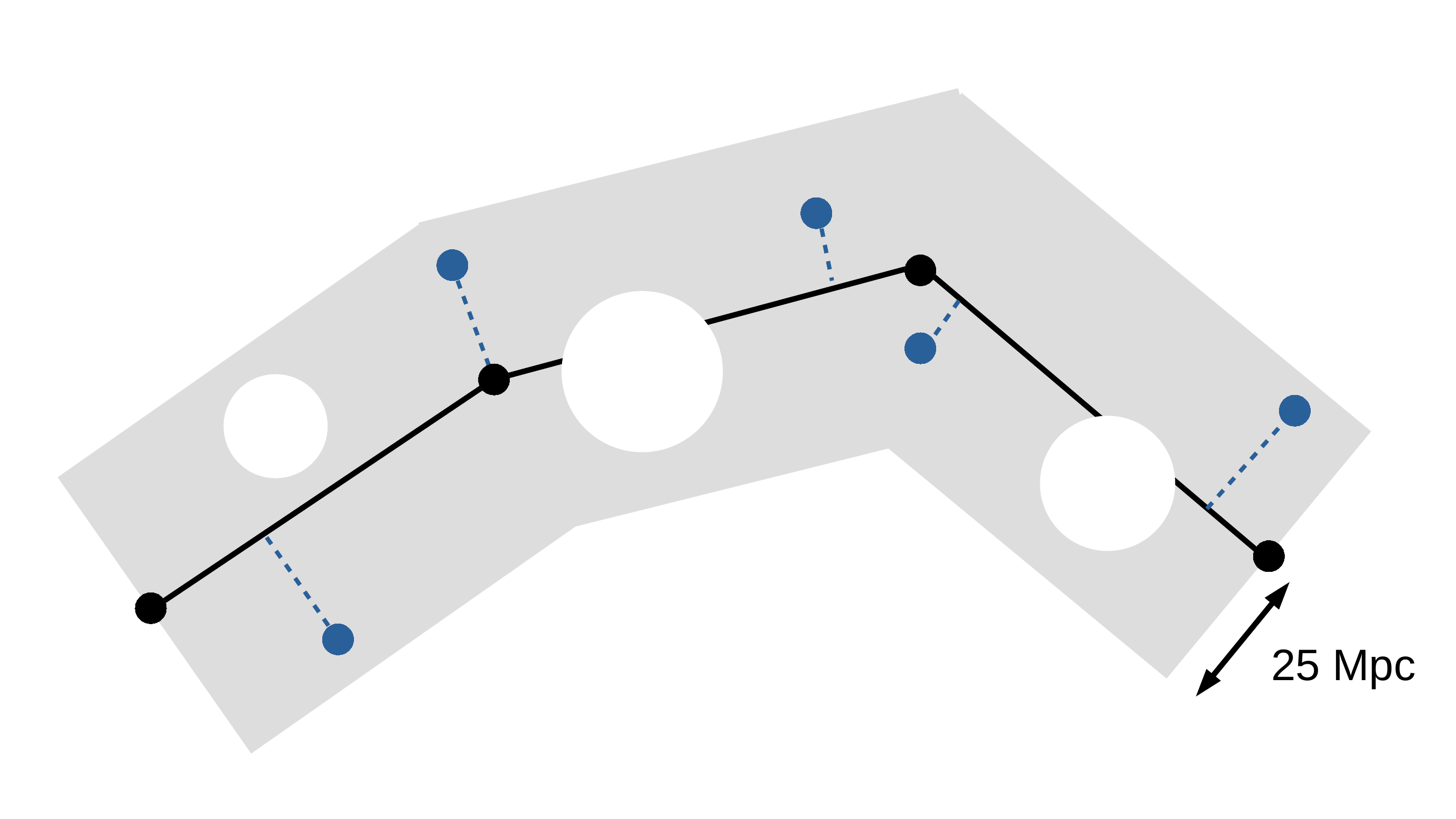}
    \caption{Steps for computing a radial filament profile. {\it Black dots} represent the extremities of a filament segment, given by DisPerSE. They are connected by {\it black straight lines}, constituting the filament spine. The {\it gray area} illustrates the area within 25 Mpc from the filament spine, where data from the tSZ or lensing map are used. For example, {\it blue points} are some data points on the tSZ or lensing map that we used to construct a radial profile. These are attached to the closest positions of the filament spine to calculate the radial distances. The radial profile is calculated based on a grid coordinate in physical distance up to 25 Mpc, divided into 20 bins (such as in Fig.~\ref{yprof}). Data in masked regions ({\it white disks}) are not used in the analysis. } 
    \label{fprof}
    \end{figure}

Our stacking procedure is described below. (i) For each filament, we extracted the region within 25 Mpc from a filament spine on the $y$ or $\kappa$ map and computed a radial profile (see Fig.~\ref{fprof}). In the figure, {\it black dots} represent the extremities of segments of a filament, given by DisPerSE. They are connected by {\it black straight lines}, which constitute the filament spine. The {\it gray area} illustrates the area within 25 Mpc from the filament spine, where data from the tSZ or lensing map are used. In this context, the {\it blue points} schematically show  the measured tSZ or lensing signal that is associated with the closest position of the filament spine and was eventually used in order to calculate the radial (tSZ or lensing) profile. The radial profile of either $y$ signal or $\kappa$ value was calculated by averaging along the filament, based on a grid coordinate in physical distance up to 25 Mpc, divided into 20 bins (such as in Fig.~\ref{yprof}). Data in masked regions ({\it white disks} in Fig.~\ref{fprof}) were not used in the analysis. In this step, one average radial profile for one filament ($y_{\rm i}(r)$ in Eq.\,\ref{eq:yprof} or $\kappa_{\rm i}(r)$ in Eq.\,\ref{eq:kprof}) was produced. 

(ii) A filament may be heavily masked when it is located close to the galactic mask or to masked clusters in the foreground or background (see Sect. \ref{subsec:ymask}). In these extreme situations, the resulting average profile becomes noisy. We removed these noisy filaments from our analysis when the resulting profile had empty bins (i.e., no data were accumulated in the bins because of the mask). This reduced the number of filaments from 27,045 to 24,544. Except for these cases, we considered masked filaments when the resulting profile was complete (no empty bin), but they were underweighted in step (iv).  

(iii) To estimate the signal excess associated with a filament, we subtracted an average signal at the outskirt (15--25 Mpc from the filament spine) as the local background signal ($y_{\rm i, bg}$ in Eq.\,\ref{eq:yprof} or $\kappa_{\rm i, bg}$ in Eq.\,\ref{eq:kprof}). In doing so, we also mitigated the contamination by foreground and background sources.

(iv) Steps (i)--(iii) were repeated for all the filaments, and then the ensemble of background-subtracted radial filament profiles were obtained. A weighted average was computed. The weights were calculated in order to minimize the effect of the mask on our measurements. For each filament, we used the ratio of unmasked area (area in {\it gray,} except for the {\it white} area in Fig.~\ref{fprof}) to the total area (area in {\it gray,} including the {\it white} area in Fig.~\ref{fprof}). The weighted average profile is given by 
\beqa
\overline{y}(r) &=& \frac{\sum w_{\rm i} \, (y_{\rm i}(r) - y_{\rm i, bg})}{\sum w_{\rm i}} \label{eq:yprof} \\
\overline{\kappa}(r) &=& \frac{\sum w_{\rm i} \, (\kappa_{\rm i}(r) - \kappa_{\rm i, bg})}{\sum w_{\rm i}} \label{eq:kprof} \\
w_{\rm i} &=& \frac{A_{\rm i, unmask}}{A_{\rm i, total}} \label{eq:weigt}, 
\eeqa
where $y_{\rm i}(r)$ or $\kappa_{\rm i}(r)$ is the $y$ or $\kappa$ radial profile of the $i$-th filament, $y_{\rm i,bg}$ or $\kappa_{\rm i,bg}$ is the average tSZ or lensing signal at 15--25 Mpc from the $i$-th filament spine, $A_{\rm i, total}$ is the total area within 25 Mpc from the $i$-th filament spine, and $A_{\rm i, unmask}$ is the unmasked area. 

\subsection{Mask for galaxy groups and clusters}
\label{subsec:ymask}
We analyzed a sample of 24,544 filaments with a length of 30--100 Mpc at $0.2<z<0.6$ ({\it red line} in Fig.~\ref{fil-z-len}). Following the stacking procedure above, we first computed the average radial $y$ or $\kappa$ profile without masking galaxy clusters ({\it blue line} in Fig.~\ref{yprof-05r500} or {\it blue line} in Fig.~\ref{kprof-05r500}). To remove the contribution from galaxy clusters, we then masked all the galaxy groups and clusters described in Sect. \ref{subsec:cluster-catalog}, with masses down to $\sim3 \times 10^{13} \, \msun$. 

However, not all the galaxy clusters are included in the catalogs, especially at high redshifts or low masses. We therefore also masked regions around critical points from DisPerSE, which are possible locations of unresolved galaxy clusters. We masked all the maxima and bifurcation points with overdensities higher than 5 in the range of 0<$z$<0.8 (the overdensity is based on the output from DisPerSE). We verified that our results did not change when we did not mask bifurcation points with lower overdensities. Saddles and minima are underdense regions and were not masked.

To investigate an appropriate mask size of galaxy clusters, we varied the radius of masked galaxy clusters and the mask size of critical points. The bright central peak associated with galaxy clusters started to disappear with an increase of mask size, and the signal amplitude stopped to decrease with the mask of $3 \times R_{500}$ for galaxy clusters and 10 arcmin for critical points. This suggests that galaxy clusters are well masked in the mask size. We therefore adopted this for our cluster mask (see details in Appendix A). 

\subsection{Significance of the tSZ signal}
\label{subsec:uncertainties}

\begin{figure}
    \centering
    \includegraphics[width=\linewidth]{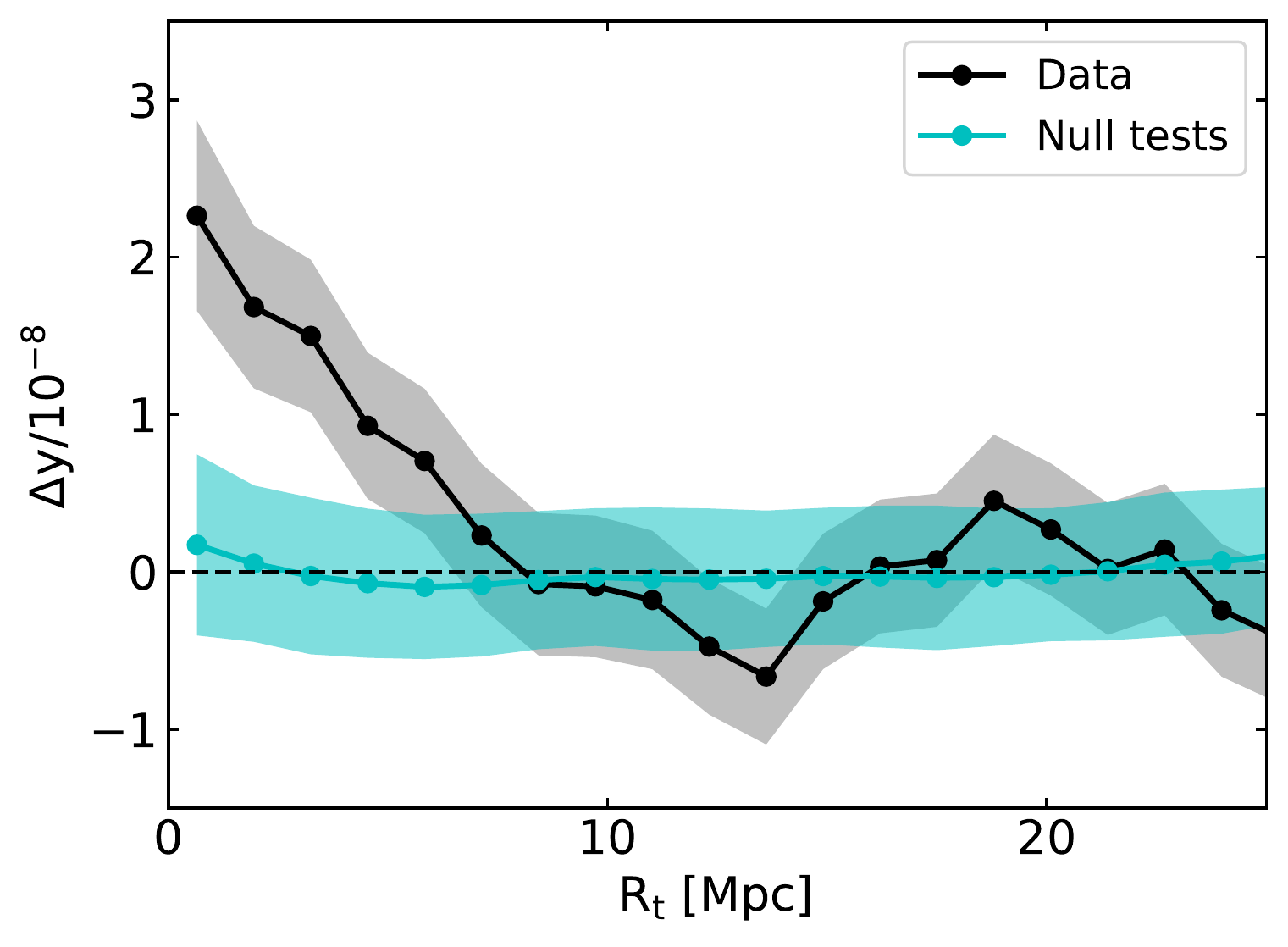}
    \caption{{\it Black}: Average radial $y$ profile of 24,544 filaments. The 1$\sigma$ uncertainty is estimated by bootstrap resampling, shown in {\it gray}. {\it Cyan}: Average radial $y$ profile from 1,000 null tests (see Sect. \ref{subsec:uncertainties}). The 1$\sigma$ uncertainty is estimated by computing a standard deviation of 1,000 null profiles. For each null test, 24,544 filaments are moved in Galactic longitude by random amounts.}
    \label{yprof}
    \end{figure}

We assessed the uncertainties of our measurements by bootstrap resampling. For this, we drew a random sampling of 24,544 filaments with replacements and recalculated the average $y$ profile for the new set of 24,544 filaments. We repeated this process 1000 times, and the bootstrapped data produced 1000 average $y$ profiles. The $rms$ fluctuation is shown in {\it gray} in Fig.~\ref{yprof}.

We also performed a null test based on a Monte Carlo method to estimate the uncertainties of the signal and assess the significance. In the null test, we moved 24,544 filaments by random angles along the Galactic longitude while keeping the galactic latitude fixed to avoid any systematic Galactic foreground or background signal, then stacked the $y$ map at the new random positions of the filaments. We repeated the null test 1000 times to determine the $rms$ fluctuations in the foreground and background sky. The average in this null-test set of profiles is consistent with zero and the null profiles have no discernible structure, as shown in ${\it cyan}$ in Fig.~\ref{yprof}. This suggests that our estimator is unbiased. In addition, the $rms$ fluctuations of the null profiles is consistent with the uncertainties from the bootstrap estimate (${\it gray}$ in Fig.~\ref{yprof}), and this ensemble of profiles can be used to estimate the uncertainties and significance of the signal.

We estimated the significance of the $y$ profile to the null hypothesis by measuring the signal-to-noise ratio (S/N) as 
\beq
S/N = \sqrt{\chi^2_{\rm data} - \chi^2_{\rm null}}
\label{eq:snr}
,\eeq
where
\beqa
\chi^2_{\rm data} &=& \sum_{i,j} y_{\rm data}(R_{i})^{T} (C^{-1}_{ij}) \, y_{\rm data}(R_{j}) \\
\chi^2_{\rm null} &=& \sum_{i,j} y_{\rm null}(R_{i})^{T} (C^{-1}_{ij}) \, y_{\rm null}(R_{j}) 
,\eeqa
where $y_{\rm data}(R_{i})$ is the $y$ value at the $R_{i}$ bin from the data $y$ profile,  $y_{\rm null}(R_{i})$ is the $y$ value at the $R_{i}$ bin from the average $y$ profile from the null tests, and $C_{ij}$ is the covariance matrix of the data $y$ profile, estimated from bootstrap resampling. When we use the data up to 10 Mpc, determined from the extent of the data profile, the S/N value is estimated to be 4.4. 

\subsection{Significance of the CMB lensing signal}
\label{subsec:kstacking}

\begin{figure}
    \centering
    \includegraphics[width=\linewidth]{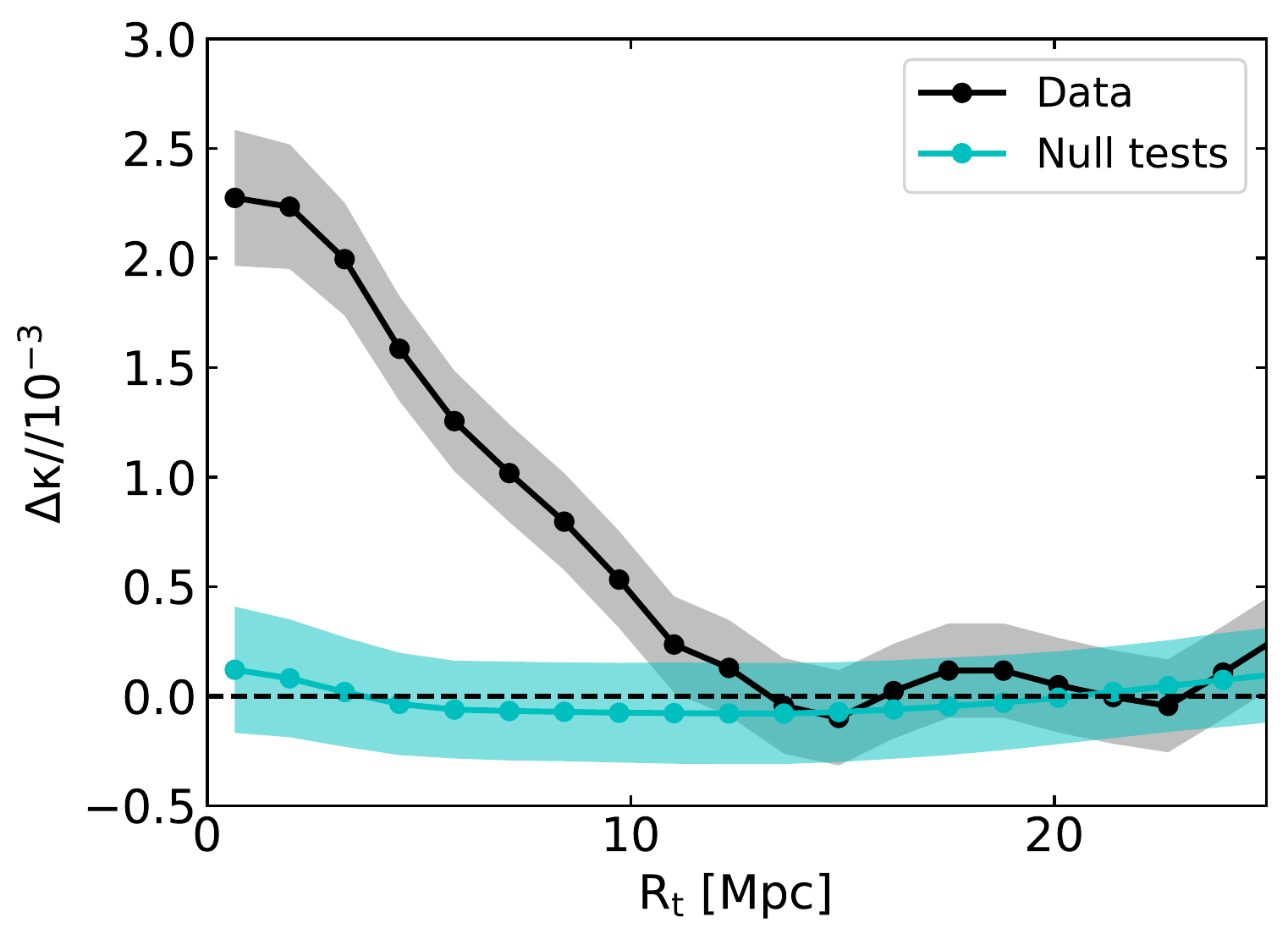}
    \caption{{\it Black}: Average radial $\kappa$ profile of 24,544 filaments. The 1$\sigma$ uncertainty is estimated by bootstrap resampling shown in {\it gray}. {\it Cyan}: Average radial $\kappa$ profile from 1,000 null tests (see Sect. \ref{subsec:uncertainties}). The 1$\sigma$ uncertainty is estimated by computing a standard deviation of 1,000 null profiles. For each null test, 24,544 filaments are moved in Galactic longitude by random amounts.}
    \label{kprof}
    \end{figure}

We repeated our stacking analysis using the \planck\ $\kappa$ map. The average $\kappa$ profile is shown in Fig.~\ref{kprof}, including the uncertainty estimate by bootstrap resampling and the result from 1000 null tests. As expected, the average of 1000 null profiles is consistent with zero, and their $rms$ fluctuations also agree with the uncertainties from the bootstrap estimate. When we use the uncertainty estimate with Eq.\,\ref{eq:snr}, the S/N of the $\kappa$ profile to null hypothesis can be estimated to be 8.1 using the data up to 10 Mpc.

\section{Systematics}
\label{sec:systematics}

In this section, we explore systematics in our tSZ and lensing measurements. Most of the discussion below applies to both types of measurements. We specify the data type when the systematics is only applied to either the tSZ or lensing measurement. 

Our measurements may be contaminated by the signal from galaxy clusters if they are not well masked by our cluster mask ($3 \times R_{500}$ for galaxy clusters and 10 arcmin for critical points). The contribution from the galaxy clusters is investigated in Appendix A by varying the radius of the masked galaxy clusters from $1 \times R_{500}$ to $4 \times R_{500}$. The bright central peak associated with galaxy clusters disappears with an increase in mask size, and the signal amplitude stops to decrease around $3 \times R_{500}$. This result suggests that the contamination from the resolved galaxy clusters is not significant in our measurements. We also masked critical points that are possible locations of unresolved galaxy groups and clusters. The fact that varying the mask sizes of critical points does not change the resulting profiles also implies that the contribution from the unresolved clusters is not significant.  

The contamination from unresolved and unmasked groups and clusters of galaxies was mostly handled by subtracting the background signals ($y$ for the tSZ and $\kappa$ for the lensing). The potential residual contamination was tested through the null tests in Sect. \ref{subsec:ystacking} for the tSZ (the result is shown in cyan in Fig.~\ref{yprof}) and in Sect. \ref{subsec:kstacking} for the lensing (the result is shown in cyan in Fig.~\ref{kprof}). As shown in the results, the average of 1000 background-subtracted profiles is consistent with zero, showing that our results are not biased by residual contamination from unresolved and unmasked clusters. The $rms$ fluctuations from the 1000 null background-subtracted profiles include the uncertainties from the unresolved and unmasked clusters, depending on their locations relative to our filament sample, and they are considered in our analysis.

Because we analyzed 3D filaments on a 2D map, the signal amplitude might be overestimated when a filament overlaps with other filaments at different redshifts. On the other hand, it might be underestimated when other filaments are located in the region of our background estimate. To investigate this effect, we simulated $y$ or $\kappa$ map by painting filament tSZ and lensing signals using our model filaments derived in Sect. \ref{sec:result}. In this procedure, filament signals at the same pixel are accumulated on the simulated map. We repeated the stacking analysis for the simulated $y$ and $\kappa$ map. The recalculated $y$ and $\kappa$ profile using the simulated $y$ and $\kappa$ map is consistent with the input profile within 10\% at the peak. This suggests that the effect of overlapping filaments is not significant for our measured profiles.

Redshift space distortions (RSD), which are not corrected for in our filament catalog, may produce some false filaments. However, false filaments are probably short and therefore are not included in our samples because of our length cut of 30 Mpc. The RSD may also change positions of filament spines from their true positions. This effect diminishes the signal amplitude and produces more extended radial profiles. This positional error can be investigated in simulations. \cite{Laigle2018} identified filaments with the DisPerSE method in the HORIZON-AGN simulations \citep{Dubois2014}: one with dark matter particles, and the other with simulated spectroscopic galaxies. The projected distances between these filaments were measured, and the authors found that the probability distribution of the distances peaks at 0.37 Mpc. This distance corresponds to 1.0 arcmin at $z$=0.47 (median redshift of our filament sample) and is negligible compared to the angular resolution of the \planck\ $y$ and $\kappa$ map. Moreover, we tested by how much the positional error of filament spines on the plane of the sky reduces the significance of our measured profile to below 3$\sigma$ (the significance of our measured profile is 4.4$\sigma$ for the tSZ and 8.1$\sigma$ for the lensing). Assuming that the positional error follows a Gaussian function, we find that a deviation of 4 Mpc suppresses our tSZ signal to below 3$\sigma$ (larger deviations are needed in the case of the lensing profile ). In the study of \cite{Laigle2018}, which was based on simulations, about 90\% of the filaments is within a distance of 4 Mpc in the cumulative distribution of distances between dark matter and spectroscopic galaxy filaments. This also suggests that the positional error of filament spines that is due to RSD is minor for our measured profiles.

The cosmic infrared background (CIB) emission in the \planck\ $y$ map might contaminate our data. This contamination may mimic our detections. The tSZ-CIB cross-correlation was carefully studied in \cite{planck2016-xxiii}. Fig.14 in the \planck\ paper shows the CIB contamination in the tSZ signal, and it is on the order of a few percent in the power spectrum on the angular scales of our filament sample of 49 -- 474 arcmin. This is not significant.


\section{Model}
\label{sec:model}

\subsection{Model of the gas distribution in filaments}
\label{subsec:gasmodel}

The Compton $y$ parameter produced by the tSZ effect is given by
\beq
y = \frac{\sigma_{\rm T} k_{\rm B}}{m_{\rm e} c^2}  \int n_{\rm e} \, T_{\rm e} \, \der l, 
\label{eq:y}
\eeq
where $\sigma_{\rm T}$ is the Thomson scattering cross section, $k_{\rm B}$ is the Boltzmann constant, $m_{\rm e}$ is the electron mass, $c$ is the speed of light, $n_{\rm e}$ is the electron number density, $T_{\rm e}$ is the electron temperature, and the integral is performed along the line-of-sight (LOS) direction. 

We can express the data $y$ profile as a geometrical projection of a density profile with $n_{\rm e}(r,z)$, 
\beq
y(R_{t},z) = \frac{\sigma_{\rm T} k_{\rm B} T_{\rm e}}{m_{\rm e} c^2}  \int_{R_{t}} \frac{2r \, n_{\rm e}(r,z)}{\sqrt{r^2 - R_{t}^2}} \, \der r, 
\label{eq:y2d}
\eeq
where $R_{t}$ is the tangential distance from a filament spine on a map. (We express the 3D distance as the lowercase letter $r$, and the 2D distance on a map as the uppercase letter $R$.) 

We can estimate the physical properties of the detected gas by considering isothermal cylindrical filaments. For the gas (electron) density profile in a filament, we considered two cases: a $\beta$ model \citep{Cavaliere1978} with $\beta = 2/3,$ and a constant density model. They are given by 
\begin{eqnarray}
n_{\rm e}(r,z) &=& \frac{n_{\rm e,0}(z)}{1+(r/r_{\rm e,c})^2} \quad \mathrm{in} \,\, r < R_{\rm max} \quad (\beta \, \rm model) \\
n_{\rm e}(r,z) &=& n_{\rm e,0}(z) \quad \mathrm{in} \,\, r < R_{\rm max} \,\, (\rm constant   \text{ }density \, model), \, 
\end{eqnarray}
where $n_{\rm e,0}(z)$ is the (central) electron density of a filament at redshift $z$, $r$ is the radius, and $r_{\rm e,c}$ is the core radius of the electron distribution, and $R_{\rm max}$ is the cutoff radius of filaments. We set $R_{\rm max}$ to be 10 Mpc for the $\beta$ model. We verified that the choice of the $R_{\rm max}$ value does not affect our results strongly. For the constant density model, $R_{\rm max}$ was set to be 5 Mpc, which was determined from the extent of the data $y$ profile. The mass distribution in filaments was found to follow $\rho \propto r^{-2}$ in N-body simulations by \cite{Colberg2005}, but the gas distribution is not well known. The validity of the model is therefore checked in Sect. \ref{subsec:fitting} by comparing the data and model profiles. 

We assumed a negligible evolution of the overdensity in filaments in the range of our sample 0.2<$z$<0.6 with a constant central electron overdensity $\delta_{\rm e,0}$. Then the electron density of a filament at redshift $z$ is given by 
\beq
n_{\rm e,0}(z) = \frac{n_{\rm e,0}(z)}{\overline{n}_{\rm e}(z)} \, \overline{n}_{\rm c}(z) = (1 + \delta_{\rm e,0}) \, \overline{n}_{\rm e}(z=0) \, (1+z)^3.
\eeq

The tSZ signal also depends on the gas temperature. We considered two models: one with constant temperature for all the filaments, and the other including temperature variations based on \cite{Gheller2019}, showing the scaling relation between baryonic mass and temperature of filaments in hydrodynamic simulations, 
\beqa
T_{\rm e}(z) &=& T_{\rm e} \quad \, (\rm constant \, temperature \, model)  \\
\label{eq:Tgas}
\mathrm{log_{10}}(T_{\rm e}(z)) &=& \alpha + \beta \, \mathrm{log_{10}}(M_{\rm BM}) \,  \\ && (\rm varied \, temperature \, model), \nonumber
\eeqa
where $\alpha$ is the normalization of the mass-temperature relation (free parameter in our analysis), $\beta$ is the power-law index of the relation, fixed to 0.315 in our analysis using the baseline model in \cite{Gheller2019}, and $M_{\rm BM}$ is the baryonic mass of filaments, calculated with Eq.\,\ref{eq:mgasfil}. In the following, our main results are described based on the constant temperature model. 

So far, the model is appropriate for a filament perpendicular to the LOS direction. We also included an orientation of the filaments that changes the amplitude of the $y$ parameter by a factor of $1/\mathrm{cos}(\theta)$ ($\theta \text{ is the}$ orientation angle of filaments from a perpendicular direction to the sky). In our study, we defined one orientation angle for one filament, using the angle between the straight line connecting both ends of a filament and the perpendicular direction to the sky. This is correct for straight filaments, but not for curved filaments. Therefore we investigated the effect of the single-angle assignment for curved filaments in our measurements. For this, we restacked the filaments after removing off-angle segments that were more than 10 degrees off from the angle defined by the straight line. By comparing the restacked profile (without off-angle segments) and the original profile (including off-angle segments), we find that their signal amplitudes are consistent within 2\%, suggesting that the single-angle assignment does not change our data interpretation. 

When the orientation angle was included, the model profile was convolved with a Gaussian kernel of 10 arcmin in FWHM, corresponding to the angular resolution of the \planck\ $y$ map, to allow a comparison with the data. It is given by 
\beq
y_{\rm mod}(R_{t},z) = \frac{ y(R_{t},z) }{ \mathrm{cos}(\theta) } * B, 
\eeq
where $y(R_{t},z$) is given by Eq.\,\ref{eq:y2d}, $\theta$ is the orientation angle of a filament defined by the angle between the straight line connecting both ends of a filament and the perpendicular direction to the sky, and $B$ is the beam, a Gaussian kernel of 10 arcmin in FWHM.

The uncertainty of the orientation angles may change our data interpretation. We therefore investigate the effect of the angular uncertainties on the Markov chain Monte Carlo (MCMC) analysis performed in Sect. \ref{subsec:fitting}. The uncertainty of the orientation angle for one filament can be estimated by calculating a standard deviation between the angle we define as a straight filament and angles of all the segments. For the ensemble of our filaments, the distribution of the angular uncertainties is described by a Gaussian function with a mean of $\text{about }$zero and a standard deviation of $\text{about }$10 degrees. To include the angular uncertainties in our MCMC analysis, we modified the orientation angle of each filament with $1/\mathrm{cos}(\theta+\delta\theta)$ ($\delta\theta$ is the uncertainty of the orientation angle for a filament). As a result, we find that the stacked model profile including the angular uncertainties is consistent with the profile that does not include them to within 7\% at the peak, and it does not significantly change our results.

\subsection{Model of the matter distribution in filaments}
\label{subsec:mattermodel}
The CMB lensing convergence is a weighted projection of matter overdensity along the LOS from the last scattering surface of the CMB  \citep{lewis2006} and is given by 
\beq
\kappa = \int^{z_{\rm CMB}}_{0} \, \der z \, W(z) \, \delta_{\rm m}(z),
\label{eq-k}
\eeq
where $z_{\rm CMB} \approx 1100$ is the redshift at the last scattering surface, $W(z)$ is the CMB lensing kernel, and $\delta_{\rm m}(z)$ is the matter overdensity at redshift $z$. For a flat universe,  $W(z)$ is given by 
\beq
W(z) = \frac{3 H_0^2 \Omega_{m,0}}{2 c H(z)} (1+z) \chi (z) \left( 1 - \frac{\chi (z)}{\chi_{\rm CMB}} \right), 
\eeq
where $H_0$ is the current Hubble parameter, $\Omega_{m,0}$ is the current matter density, $\chi (z) $ is the comoving distance at redshift $z,$ and $\chi_{\rm CMB}$ is the comoving distance to the last scattering surface. 

To model the $\kappa$ profile in Fig.~\ref{kprof}, we followed Sect. \ref{subsec:gasmodel}. We considered a cylindrical filament, and for the matter density profile, we also considered the $\beta$ model with $\beta = 2/3$ and a constant density model. They are given by 
\begin{eqnarray}
\delta_{\rm m}(r) &=& \frac{\delta_{\rm m, 0}}{1+(r/r_{\rm m,c})^2} \quad \mathrm{in} \,\, r < R_{\rm max} \quad (\beta \, \rm model) \\
\delta_{\rm m}(r) &=& \delta_{\rm m, 0} \quad \mathrm{in} \,\, r < R_{\rm max} \quad (\rm constant \, density \, model), \, 
\end{eqnarray}
where $\delta_{\rm m,0}$ is the (central) matter overdensity of filaments, which is constant in our model, $r_{\rm m,c}$ is the core radius of matter distribution, and $R_{\rm max}$ is 10 Mpc for the $\beta$ model and 5 Mpc for the constant density model. This 3D matter density profile is projected onto 2D, including an orientation angle, and its amplitude is suppressed at $\ell > 400$ ($r\sim$ 6 Mpc) with an exponential function in the same manner as we applied to the \planck\ CMB lensing map (see Sect. \ref{subsec:kmap}).


\section{Results}
\label{sec:result}


\subsection{MCMC}
\label{subsec:mcmc}
We fit our model to the data by computing a minimum chi-square value. The fitting was performed for the $y$ and $\kappa$ profiles simultaneously by assuming that gas follows dark matter in filaments: $\delta_{\rm e,0} \simeq \delta_{\rm m,0} \simeq \delta$ and $r_{\rm e,c} \simeq r_{\rm m,c} \simeq r_{\rm c}$. This assumption is supported by \cite{Gheller2019}, who showed that baryon fractions in filaments are almost equivalent to the cosmic baryon fraction ($\sim$0.9 of the cosmic baryon fraction) independent of baryonic effects such as stellar and AGN feedback using their hydrodynamic simulations. Under this assumption, our estimator for the fitting is 
\beqa
\chi^2 &=& \chi^2_{y} + \chi^2_{\kappa} \\
\chi^2_{y} &=& \sum_{i,j} (y(R_{i}) - y_{\rm mod}(R_{i}))^{T} (C^{-1}_{y,ij}) \, (y(R_{j}) - y_{\rm mod}(R_{j})) \\
\chi^2_{\kappa} &=& \sum_{i,j} (\kappa(R_{i}) - \kappa_{\rm mod}(R_{i}))^{T} (C^{-1}_{\kappa, ij}) \, (\kappa(R_{j}) - \kappa_{\rm mod}(R_{j})), 
\label{eq:chi2}
\eeqa
where $y (R_{i})$ and $\kappa (R_{i})$ are the $y$ and $\kappa$ values at $i$-th bin in the data, and $y_{\rm mod} (R_{i})$ and $\kappa_{\rm mod} (R_{i})$ are the corresponding values in the model. $C_{y,ij}$ and $C_{\kappa,ij}$ is the covariance matrix of the data $y$ and $\kappa$ profile, estimated by the bootstrap resampling. The fitting was performed with the MCMC algorithm using the emcee software \citep{foreman2013}, which is an affine-invariant ensemble sampler proposed by \cite{goodman2010}. 

\subsection{Model fitting}
\label{subsec:fitting}
\begin{figure}
    \centering
    \includegraphics[width=\linewidth]{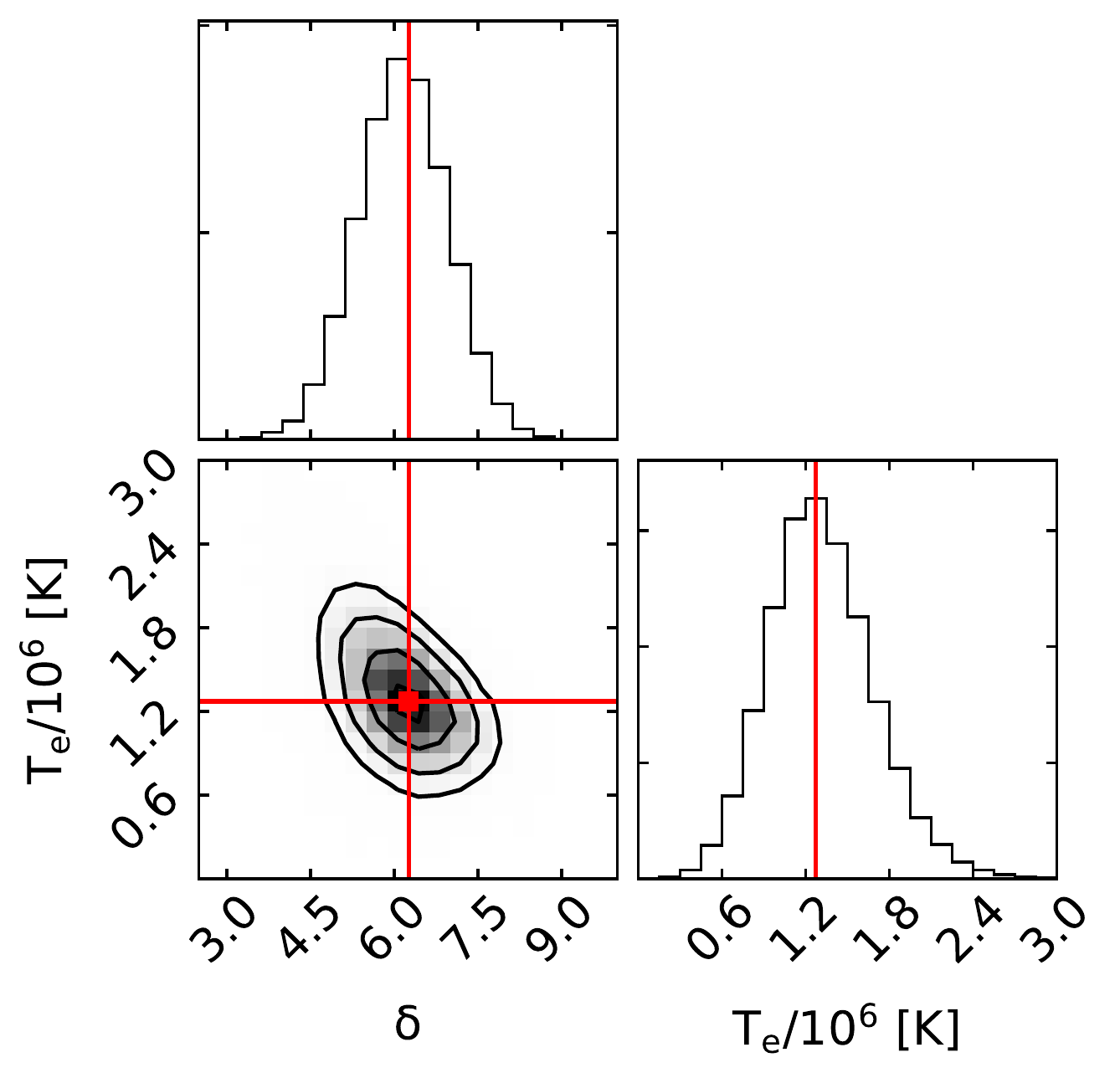}   
    \caption{Probability distributions of gas and matter overdensity $\delta$ and gas temperature $T_{\rm e}$ for the constant density and constant temperature model, obtained by the MCMC sampling. The panels in the diagonal show the 1D histogram for each model parameter obtained by marginalizing over the other parameters, with a red line to indicate the best-fit value. The off-diagonal panels show 2D projections of the posterior probability distributions for each pair of parameters, with contours to indicate 1$\sigma$, 2$\sigma,$ and 3$\sigma$ regions.}
    \label{ovTe-mcmc}
    \end{figure}

We fit two free parameters in the constant density and constant temperature models (gas and matter overdensity $\delta$ and gas temperature $T_{\rm e}$) with the MCMC. The result of the MCMC is shown in Fig.~\ref{ovTe-mcmc}, providing the best-fit values of $\delta = 6.3^{+0.9}_{-0.8}$ and $T_{\rm e} = 1.3^{+0.4}_{-0.4} \times 10^6$ [K]. An anticorrelation is seen between the overdensity and temperature that originates in the tSZ, and the lensing data help to constrain the overdensity (the lensing data alone constrain the overdensity).

\begin{figure}
    \centering
    \includegraphics[width=\linewidth]{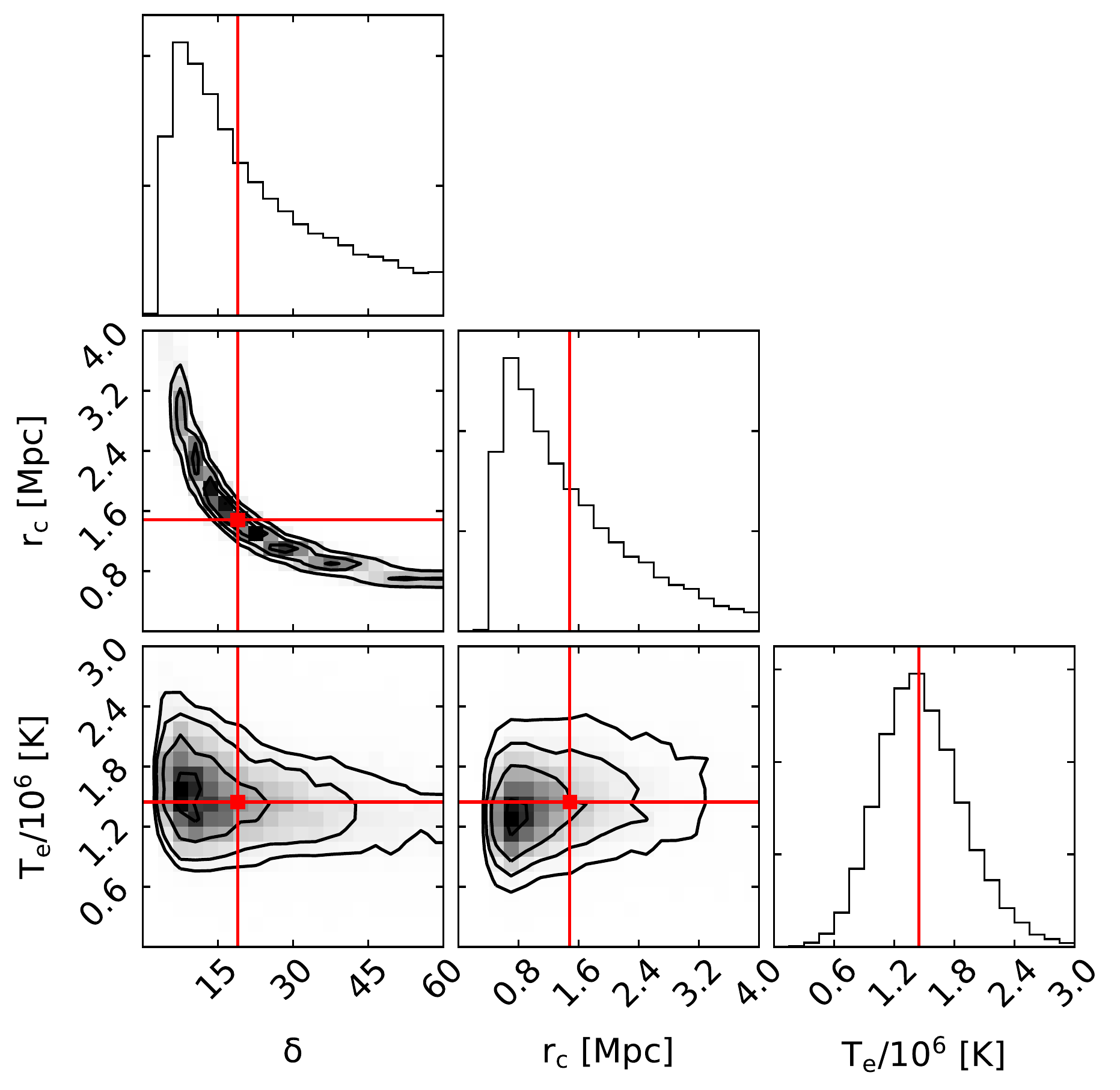}   
    \caption{Probability distributions of gas and matter overdensity $\delta$, core radius $r_{\rm c}$ , and gas temperature $T_{\rm e}$ for the $\beta$ model and constant temperature model, obtained by the MCMC sampling. The panels in the diagonal show the 1D histogram for each model parameter obtained by marginalizing over the other parameters, with a red line to indicate the best-fit value. The off-diagonal panels show 2D projections of the posterior probability distributions for each pair of parameters, with contours to indicate 1$\sigma$, 2$\sigma,$ and 3$\sigma$ regions. } 
    \label{ovrcTe-mcmc}
    \end{figure}

The MCMC fitting was also performed for the $\beta$ model and constant temperature model with three free parameters: core radius $r_{\rm c}$, central gas and matter overdensity $\delta,$ and gas temperature $T_{\rm e}$. We added uniform priors, $0 < \delta < 100$, $0 < r_{\rm c} < 5$ [Mpc], and  $10^5 < T_{\rm e} < 10^7$ [K], which are the ranges predicted by hydrodynamic simulations. With these priors, the result of the MCMC is shown in Fig.~\ref{ovrcTe-mcmc}, providing a best-fit values of $\delta = 19.0^{+27.3}_{-12.1}$, $r_{\rm c} = 1.5^{+1.8}_{-0.7}$ [Mpc], and $T_{\rm e} = 1.4^{+0.4}_{-0.4} \times 10^6$ [K]. The overdensity and core radius have relatively large uncertainties because they are strongly degenerate. This degeneracy is difficult to break with the current data set, but can be broken by future CMB measurements with better sensitivity and angular resolution. 

\begin{figure}
    \centering
    \includegraphics[width=\linewidth]{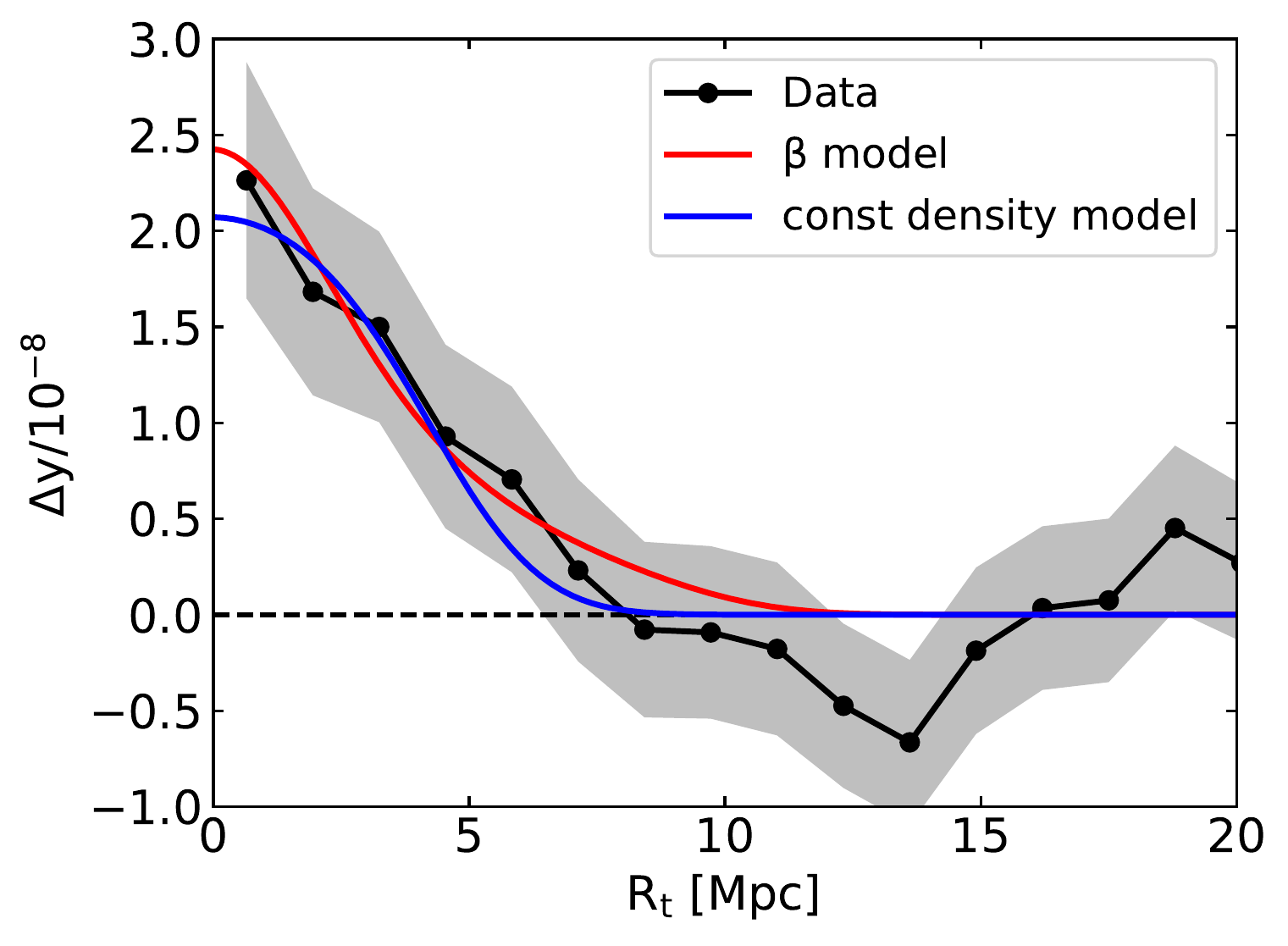}
    \includegraphics[width=\linewidth]{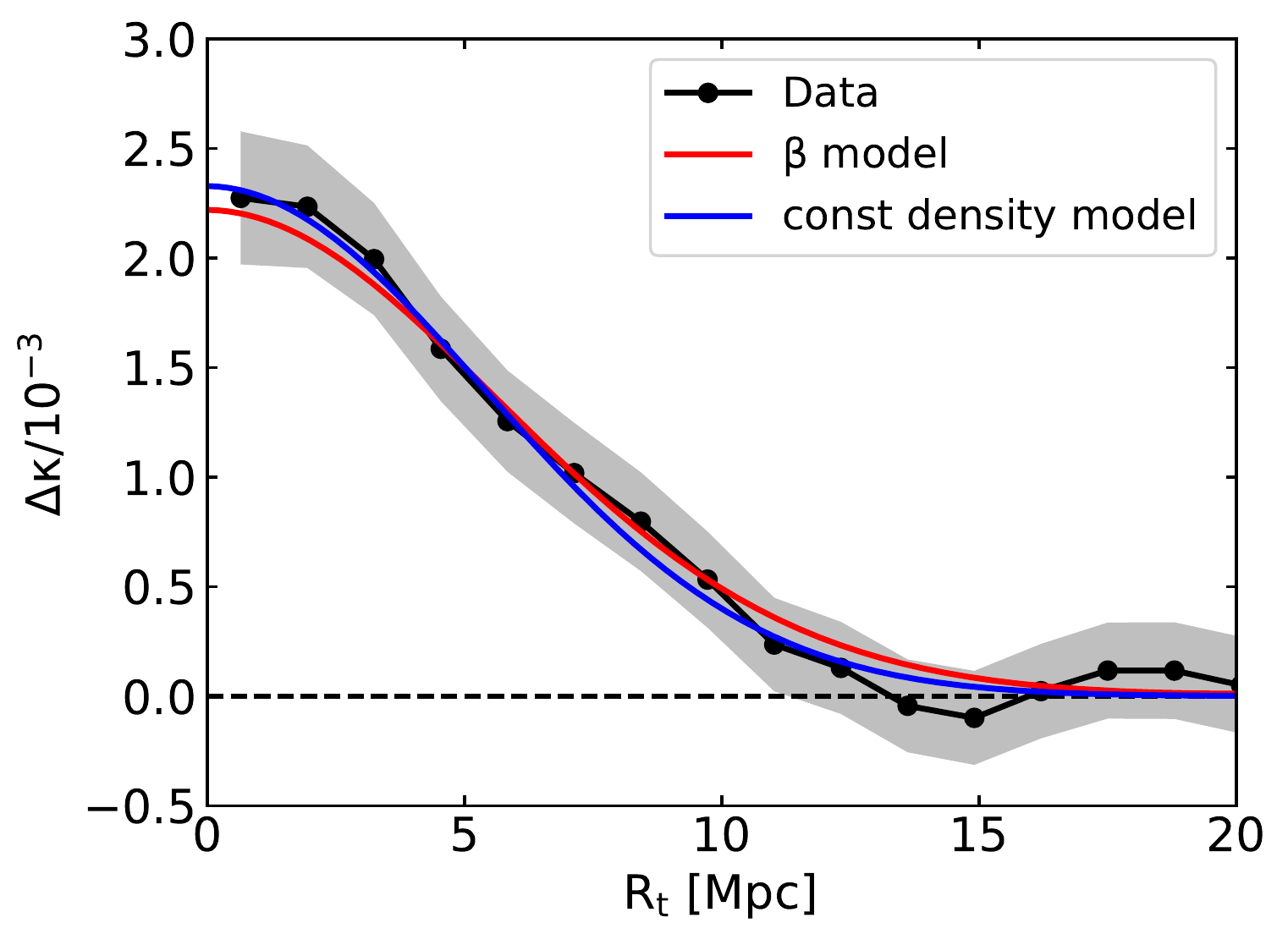}
    \caption{Average radial $y$ ({\it upper}) and $\kappa$ ({\it lower}) profile of 24,544 filaments, fit with the $\beta$ model ({\it red}) and constant density model ({\it blue}) along with the constant temperature model (see Sect. \ref{sec:model} for the models). }
    \label{ykprof-fit}
    \end{figure}

Figure~\ref{ykprof-fit} shows the model profiles with the best-fit values derived above. The reduced $\chi^2$ values are 0.6 for the $\beta$ model (16 data points and three fit parameters) and 0.6 for the constant density model (16 data points and two fit parameters). Both models fit the data quite well, and no significant difference between these models is found. 

\subsection{Baryon budget in filaments}
\label{subsec:baryonbudget}

We estimated the contribution of our measurements to the cosmic baryons assuming that gas (electron) follows dark matter ($\delta_{\rm e} \simeq \delta_{\rm m}$), as discussed in Sect. \ref{subsec:mcmc}. The gas-mass density can be calculated from an electron overdensity by 
\beq
\rho_{\rm gas}(z) = \overline{n}_{\rm e}(z) (1 + \delta_{\rm e}) \, \mu_{\rm e} \, m_{\rm p}, 
\eeq
where $\overline{n}_{\rm e}(z)$ is the mean electron number density in the Universe at redshift $z$, $\mu_{\rm e} = \frac{2}{1+\chi} \simeq 1.14$ is the mean molecular weight per free electron for a cosmic hydrogen abundance of $\chi = 0.76$,  and $m_{\rm p}$ is the mass of the proton. With the gas-mass density, the total gas mass in a filament at $z$ is given by 
\beqa
M_{\rm gas,fil}(z) &=& \int_0^{L_{\rm fil}} \der l \int_0^{R_{\rm max}} \der r \, 2 \pi r \, \rho_{\rm gas}(r,z), \label{eq:mgasfil} \\
\rho_{\rm gas}(r,z) &=& n_{\rm e}(r,z) \, \mu_{\rm e} \, m_{\rm p} 
\eeqa
where $L_{\rm fil}$ is the length of a filament, and $R_{\rm max}$ is the size of the filament (See Sect. \ref{sec:model}). The gas masses of the filaments used in our analysis range from $1.7 \times 10^{14} \, \msun $ to $1.3 \times 10^{15} \, \msun$ for the density distribution from the $\beta$ model. We note that the gas masses are estimated to be higher by $\sim$8\% for the constant density model. By summing the gas masses of the filaments, the gas mass density can be calculated by 
\beq
\rho_{\rm gas,fil} = \frac{\sum_{i}^{N_{\rm fil}} M_{\rm gas,fil,i}}{V_{\rm c} \times f_{\rm SDSS}},  
\eeq
where $V_{\rm c}$ is the comoving volume, between $z$ = 0.2 and 0.6 in our case (redshift range of our filament sample), and $f_{\rm SDSS} \simeq 0.18$ is the fractional SDSS-DR12 survey field on the sky (with no mask). By summing the gas masses in 24,544 filaments used in our analysis (30--100 Mpc at $0.2 < z < 0.6$), the gas mass density of the filaments relative to the cosmic baryon density is estimated to be $\rho_{\rm gas,fil}/\overline{\rho}_{\rm b} = 0.080^{+0.116}_{-0.051}$ for the $\beta$ model and $\rho_{\rm gas,fil}/\overline{\rho}_{\rm b} = 0.087^{+0.012}_{-0.010}$ for the constant density model, respectively. 

\subsection{MCMC analysis with a varied temperature model}
\label{subsec:variedtemp}
We also performed the MCMC analysis with a varied temperature model, which was combined with the constant-density model and the $\beta$ model. The MCMC results (best-fit overdensity, core radius, and baryon budget, and their uncertainties) are almost same as those derived with the constant temperature model. The varied temperature model gives an additional temperature variation depending on the gas mass in filaments (see Eq.\,\ref{eq:Tgas}). The gas-mass range is estimated in Sect. \ref{subsec:baryonbudget} and corresponds to $(0.9 - 1.8) \times 10^{6}$ K with the constant density model and $(1.0 - 1.9) \times 10^{6}$ K with the $\beta$ model. 


\section{Discussion}
\label{sec:discussion}

We estimate that the overdensity at the core of the filaments with the $\beta$ model is $\delta\sim19$. The value is well within the expected values by hydrodynamic simulations ($10<\delta<100$) (\ie, \citealt{Cen1999, cen2006, Martizzi2018}). 

Other studies have statistically investigated gas and matter of relatively short filaments and found them to be on the order of $\sim$10 \mpc\ . They mostly show the average overdensity inside filaments, therefore we compare these results to our estimated overdensity with the constant density model ($\delta\sim6.3$) as follows.

\cite{Epps2017} have examined the weak-lensing signal between $\text{about }$ 23,000 pairs of SDSS-III and BOSS LRGs with tangential distances between 6 \mpc\ and 10 \mpc\ using the  Canada-France-Hawaii telescope lensing survey (CFHTLenS) data. They found that the stacked filaments have an average mass of $(1.6 \pm 0.3) \times 10^{13} \, \msun$ in a region of 7.1 $h^{-1} \mathrm{Mpc}$ long and 2.5 $h^{-1} \mathrm{Mpc}$ wide and estimated an overdensity of $\delta \sim 4$ in the filaments assuming a uniform density cylinder. 

Similar studies were performed for gas in filaments using the \planck\ tSZ data in \cite{Tanimura2019b} and \cite{deGraaff2019}. \cite{Tanimura2019b} studied filaments between 260,000 SDSS-LOWZ LRG pairs at $z \sim$0.3 with tangential distances between 6 \mpc\ and 10 \mpc\ and estimated the overdensity of gas to be $\delta \sim 3.2$ using the temperature estimate from hydrodynamic simulations. Likewise, \cite{deGraaff2019} studied filaments between about one million SDSS-CMASS galaxy pairs at $z \sim$0.55 with tangential distances between 6 \mpc\ and 14 \mpc\ and estimated the overdensity of gas to be $\delta \sim 5.5$ along with the \planck\ CMB lensing data. 

These density estimates ($\delta = 3.2\sim$5.5) are relatively consistent with our estimated gas and matter overdensity in filaments assuming a uniform density, $\delta \sim$6.3. Our estimated value is slightly higher than that of others. The reason may be that previous analyses assumed that pairs of LRGs are always connected by filaments, but some stacked pairs are not. On the other hand, we know the filament locations. The overdensity value can also be compared with \cite{Cautun2014}, who reported a probability distribution of overdensities in filaments in the Millennium simulations. The distribution peaks around $\delta \sim 3$, while it spreads over three orders of magnitude from $\sim$0.1 to $\sim$100. 

The fact that the average overdensities in filaments on scales of $\sim$10 Mpc and $\sim$50 Mpc (median length of our filaments) are relatively consistent may imply a universal density distribution in filaments. The universality may be supported by the studies of hydrodynamic simulations in \cite{Gheller2019}, who reported that baryon fractions in filaments are almost equivalent to the cosmic baryon fraction ($\sim$0.9 of the cosmic baryon fraction) independent of the baryonic effects that are included in the simulations, such as stellar and AGN feedback. However, their identification of filaments is relatively simple, and a more dedicated study is still needed.

Our measured baryons in 24,544 filaments (30--100 Mpc) are estimated to be $0.080^{+0.116}_{-0.051} \times \Omega_{\rm b}$. The value can increase to $0.141 \times \Omega_{\rm b}$ when gas masses of shorter (< 30 Mpc) and longer filaments (> 100 Mpc) in the filament catalog are added, considering that the gas density in filaments may be universal independent of their lengths, as discussed above. The gas masses of the shorter and longer filaments were calculated using Eq.\,\ref{eq:mgasfil}, in which the $\beta$ model is used for the density distribution. Our measured baryon fraction is consistent with that of  \cite{deGraaff2019} of $(0.11 \pm 0.07) \times \Omega_{\rm b}$. This result is expected because we used the same dataset: SDSS-CMASS galaxies and \planck\ tSZ and lensing map, which implies that we trace the same WHIM gas. 

Gas temperatures in filaments were measured with relatively large uncertainties. For example, the gas temperature in the northeast filament of A2744 is estimated to be $0.27^{+0.09}_{-0.05}$ keV ($\sim 3.1 \times 10^{6}$ K) by \suzaku\ observations in \cite{hattori2017}. In addition, the gas temperature in filaments between about one million CMASS galaxy pairs with distances between 6 and 14 \mpc\ is estimated to be $(2.7 \pm 1.7) \times 10^{6}$ K \citep{deGraaff2019} by combining the tSZ and CMB lensing measurements. These temperature estimates are slightly higher than our estimate of $T_{\rm e} = 1.4^{+0.4}_{-0.4} \times 10^6$ [K]. One reason may be that their filaments are shorter and influenced thermally by nearby galaxy groups and clusters. Another reason may be that we masked the resolved galaxy groups and clusters and critical points in the density field described in Sect. \ref{subsec:ymask} to minimize the contamination from the foreground and background galaxy groups and clusters, which was not done in their studies.

\section{Conclusion}
\label{sec:conclusion}

We have studied gas and matter in the filaments identified by the DisPerSE method using the \planck\  Sunyaev-Zel’dovich (tSZ) map and \planck\ CMB lensing map.  To remove the contribution from galaxy clusters, we masked all the galaxy groups and clusters detected by the \planck\ SZ, \rosat\ X-ray and SDSS optical surveys down to the total mass of $\sim3 \times 10^{13} \, \msun$ , and also critical points identified by DisPerSE that are possible locations of resolved or unresolved galaxy groups and clusters. After masking them, we stacked the \planck\ $y$ map at the positions of the filaments and detected the tSZ signal at a significance of 4.4$\sigma$. The detected signal extends out to 8 Mpc from the filament spines. We applied the same stacking procedure for the \planck\ $\kappa$ map and detected the lensing signal at 8.1$\sigma$ significance.

We fit the radial profiles of filaments measured by the tSZ and lensing assuming that gas (electron) follows dark matter, $\delta_{\rm e} \simeq \delta_{\rm m}$. When we consider isothermal cylindrical filaments with the $\beta$ model ($\beta$=2/3) for the gas and matter density profile, the central overdensity of gas and matter can be estimated to be $\delta = 19.0^{+27.3}_{-12.1}$ with the core radius of $r_{\rm c} = 1.5^{+1.8}_{-0.7}$ [Mpc]. With these values, the $\beta$ model fits the $y$ and $\kappa$ radial profiles well. By replacing the density model from the $\beta$ model to the constant density model, the overdensity estimate becomes $\delta = 6.3^{+0.9}_{-0.8}$. We also estimate the gas (electron) temperature in the filaments. Assuming a constant temperature, it is found to be $T_{\rm e} = (1.4 \pm 0.4) \times 10^6$ [K] with the $\beta$ model for the density and $T_{\rm e} = (1.3 \pm 0.4) \times 10^6$ [K] with the constant density model. Under the assumption of a varying temperature, we find that it ranges from $\sim1.0 \times 10^6$ [K] to $\sim1.9 \times 10^6$ [K]. 

Furthermore, we estimate the total amount of baryons measured in the filaments. It amount to $0.080^{+0.116}_{-0.051} \times \Omega_{\rm b}$ in 24,544 filaments with 30--100 Mpc long and may increase to $0.141 \times \Omega_{\rm b}$ by adding gas masses of shorter (< 30 Mpc) and longer filaments (> 100 Mpc) to the filament catalog, although we did not use them in our analysis, assuming that the gas density in filaments is universally independent of their lengths. This assumption can be inferred from the fact that our density estimate using filaments with $\sim$50 Mpc ($\delta \sim$6.3) is relatively consistent with the values derived from other measurements for filaments with $\sim$10 \mpc\ ($\delta = 3.2\sim$5.5). 

Our study can be extended with larger spectroscopic surveys such as the extended BOSS (eBOSS) in SDSS-IV and the Dark Energy Spectroscopic Instrument (DESI) \citep{DESI2016}, as well as future missions such as Large Synoptic Survey Telescope (LSST) \citep{lsst2008, lsst2009} and Euclid \citep{Laureijs2011}. Their larger samples will play an important role in the more precise identification of the cosmic-web structure. In addition, their deeper observations will shed light on the evolution of the cosmic-web structure with redshift and enable us to test cosmological models. Moreover, the distribution and physical state of gas in the cosmic web may be further studied statistically by \erosita\ X-ray observations \citep{Merloni2012} as well as through more sensitive future CMB measurements with high angular resolution such as Cosmic Microwave Background Stage 4 (CMB-S4) \citep{Abazajian2016}.


\appendix

\section{Masking galaxy clusters}
\label{sec:masktest}
Our measurements may be contaminated by the signal from massive galaxy clusters at the foreground or background if they are not well masked by our cluster mask ($3 \times R_{500}$ for galaxy clusters and 10 arcmin for critical points). As a reminder, we masked critical points of all maxima and bifurcations with overdensities higher than 5, and saddles and minima are underdense regions and were not masked (see Sect. \ref{subsec:ymask}). We investigated the contribution from galaxy clusters by varying the radius of the masked galaxy clusters from $0 \times R_{500}$ to $4 \times R_{500}$ and the mask size of critical points from 0 arcmin to 15 arcmin.

\begin{figure}[ht!]
    \centering
    \includegraphics[width=\linewidth]{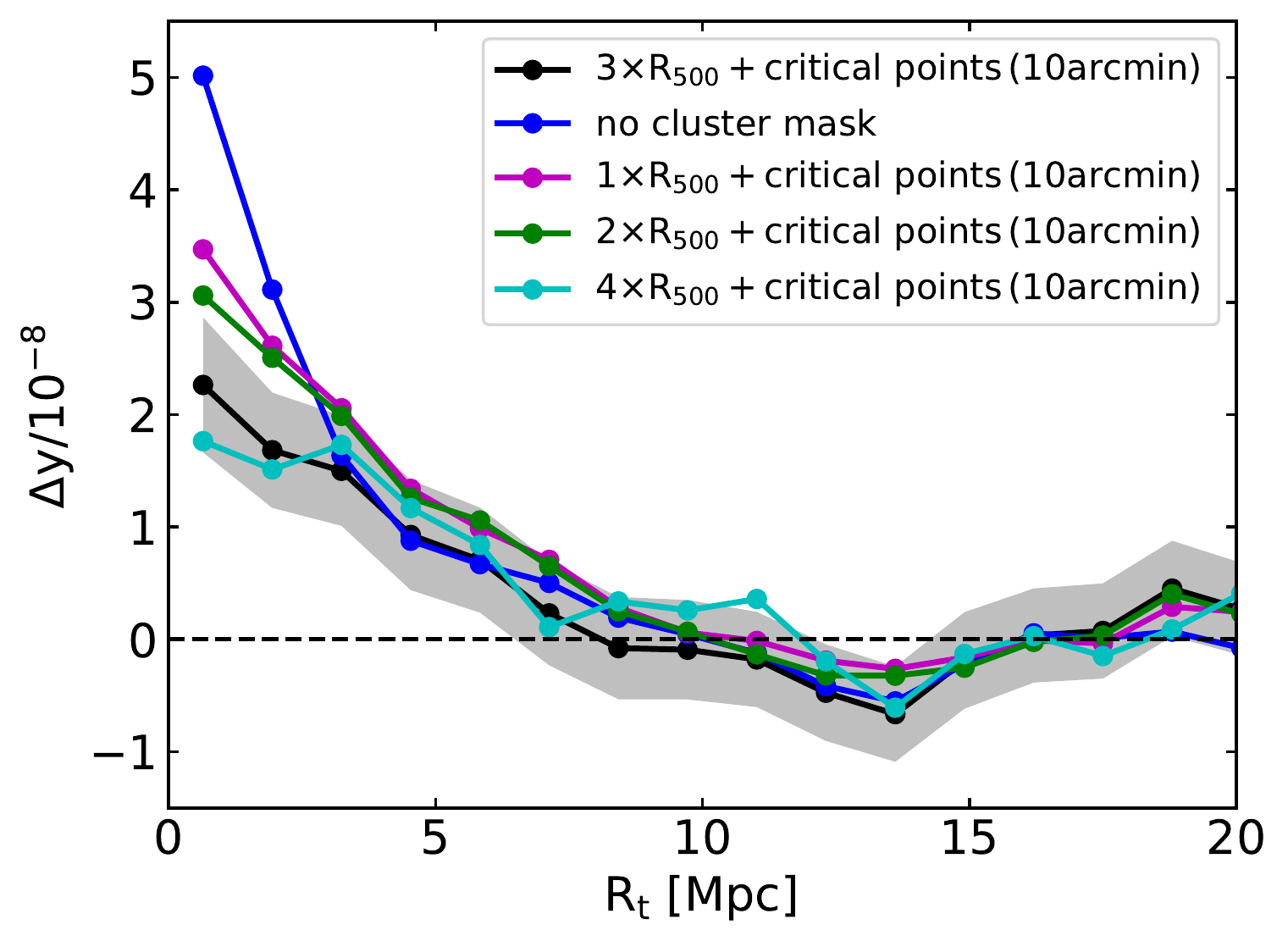}
    \includegraphics[width=\linewidth]{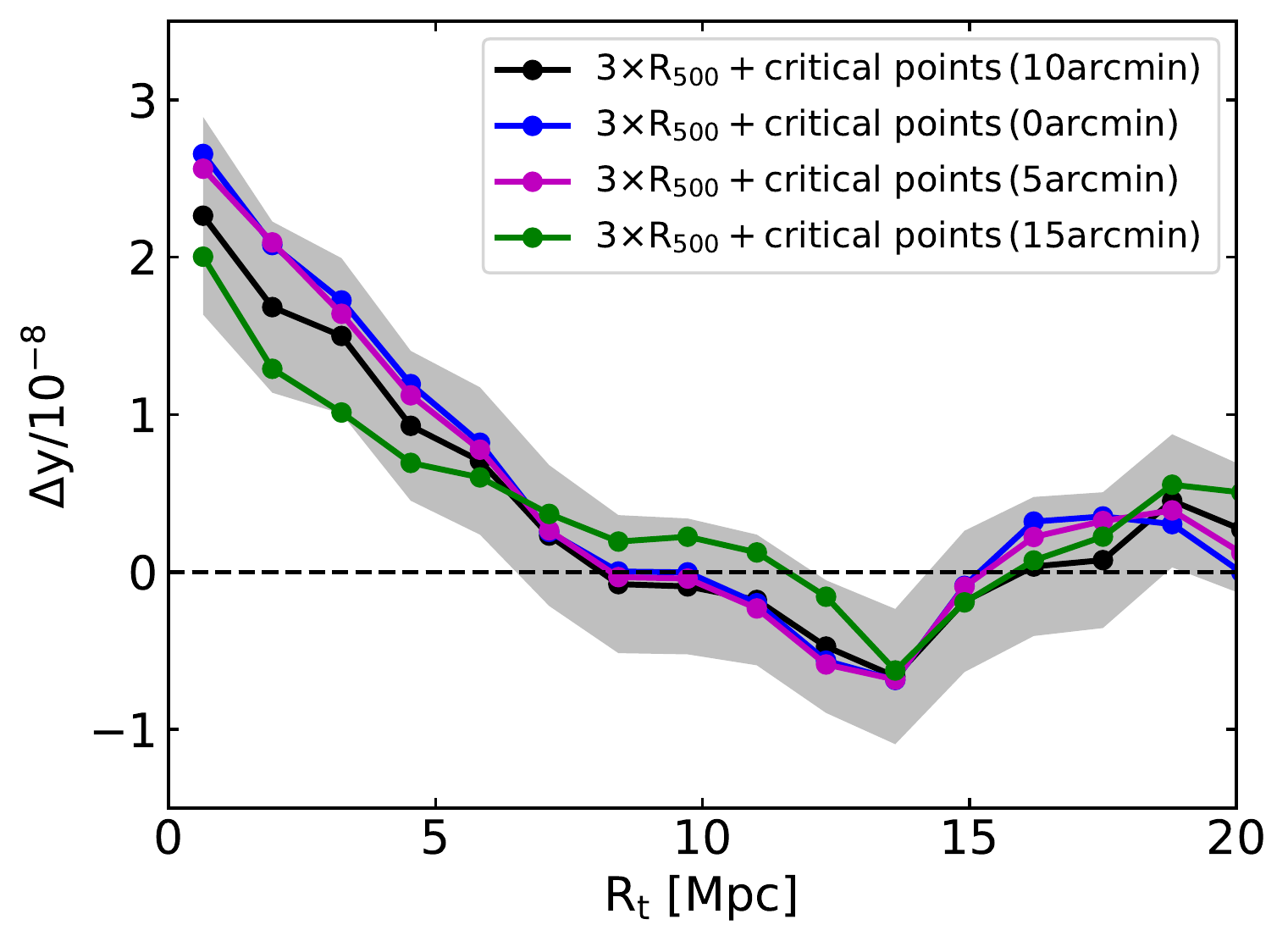}
    \caption{{\it Upper}: Average radial $y$ profiles of 24,544 filaments with different radii of the cluster masks: No cluster mask ({\it blue}), $1 \times R_{500}$ ({\it magenta}), $2 \times R_{500}$ ({\it green}), $3 \times R_{500}$ ({\it black}), and $4 \times R_{500}$ ({\it cyan}). {\it Lower}: Average radial $y$ profiles of 24,544 filaments with different sizes of critical point masks: No critical point mask ({\it blue}), 5 arcmin mask ({\it magenta}), 10 arcmin mask ({\it black}), and 15 arcmin mask ({\it green}). In both figures, the 1$\sigma$ uncertainties of the profile with ($3 \times R_{500}$ + 10 arcmin) mask are estimated by bootstrap resampling and are shown in {\it gray} (see Sect. \ref{subsec:uncertainties}). }
    \label{yprof-05r500}
    \end{figure}

First, we varied the radius of masked galaxy clusters, while the mask size of critical points was fixed at 10 arcmin. We show this in the {\it upper panel} of Fig.~\ref{yprof-05r500}. The bright central peak associated with galaxy clusters disappears with an increase in mask size, and the signal amplitude stops to decrease at $3 \times R_{500}$. 

We also varied the mask size of critical points while the mask size of galaxy clusters was fixed at $3 \times R_{500}$ in the {\it lower panel} of Fig.~\ref{yprof-05r500}. The resulting profiles are consistent because resolved massive clusters are already masked by the $3 \times R_{500}$ mask. Masking of critical points may not be necessary because they are most likely unresolved low-mass clusters. However, we see a weak signal excess when we stack the $y$ map at the positions of critical points, therefore we masked the critical points in a radius of 10 arcmin, which is the angular resolution of the \planck\ $y$ map.

\begin{figure}
    \centering
    \includegraphics[width=\linewidth]{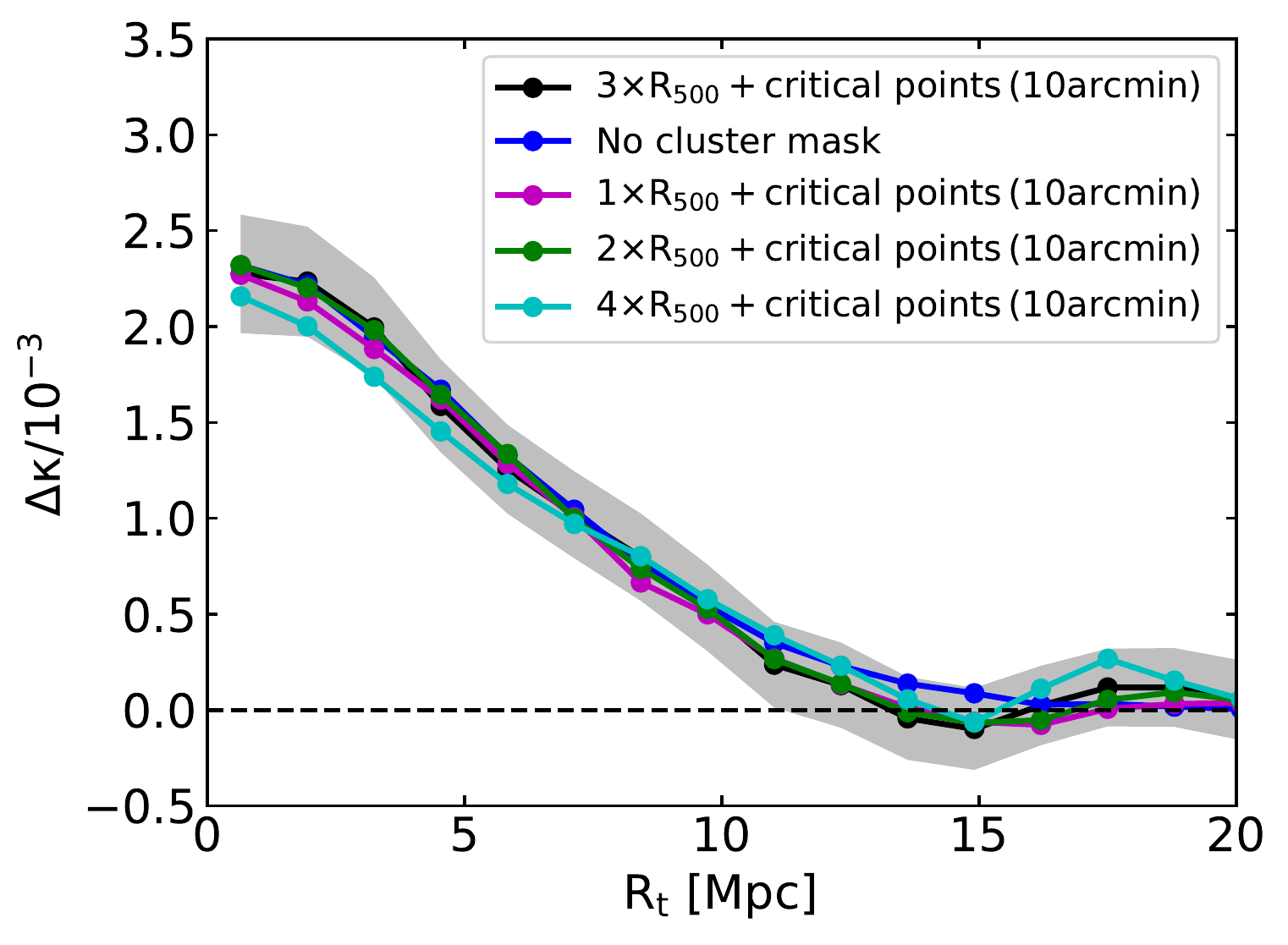}
    \includegraphics[width=\linewidth]{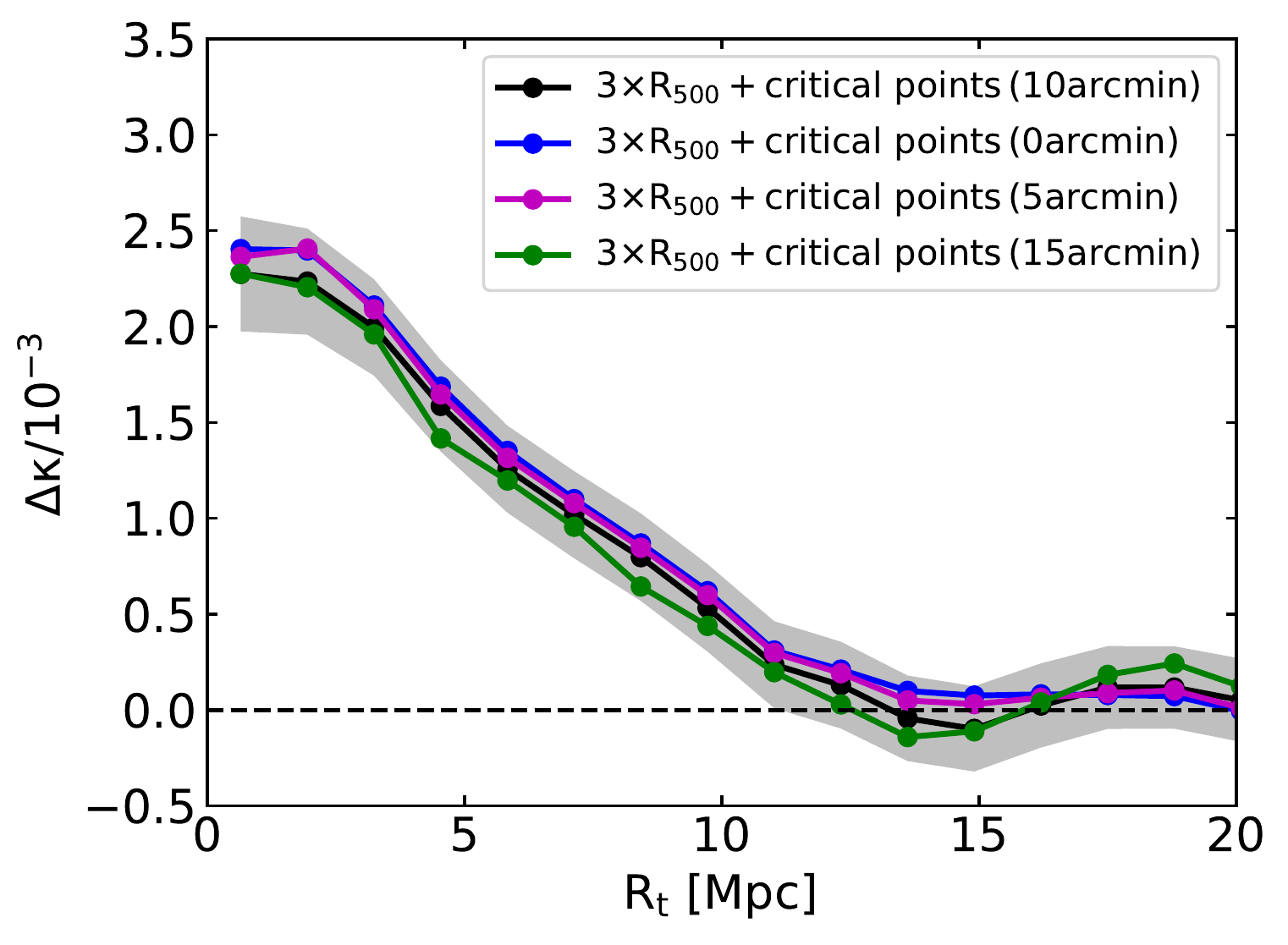}
    \caption{{\it Upper}: Average radial $\kappa$ profiles of 24,544 filaments with different radii of the cluster masks: No cluster mask ({\it blue}), $1 \times R_{500}$ ({\it magenta}), $2 \times R_{500}$ ({\it green}), $3 \times R_{500}$ ({\it black}), and $4 \times R_{500}$ ({\it cyan}). {\it Lower}: Average radial $\kappa$ profiles of 24,544 filaments with different sizes of critical point masks: No critical point mask ({\it blue}), 5 arcmin mask ({\it magenta}), 10 arcmin mask ({\it black}), and 15 arcmin mask ({\it green}). In both figures, the 1$\sigma$ uncertainties of the profile with ($3 \times R_{500}$ + 10 arcmin) mask are estimated by bootstrap resampling and are shown in {\it gray} (see Sect. \ref{subsec:uncertainties}). }
    \label{kprof-05r500}
    \end{figure}

We also investigated the contamination from massive galaxy clusters in our lensing measurements by varying the radius of masked galaxy clusters and the mask size of critical points in Fig.~\ref{kprof-05r500}. The figures show that the resulting profiles are all consistent, and we failed to find the decreasing trend with increasing mask radius seen in the tSZ (Fig.~\ref{yprof-05r500}). This result is most likely due to the fact that the \planck\ SZ clusters are already masked in the original mask for the \planck\ $\kappa$ map (see Sect. \ref{subsec:kmap}). It also suggests that the contribution from lower-mass clusters is minor in our lensing measurements. However, we adopted the same cluster mask as we used in the tSZ analysis ($3 \times R_{500}$ for galaxy clusters and 10 arcmin for critical points) in the lensing analysis as well for an equivalent comparison of gas and matter measurements using the same sky area.  

\begin{acknowledgements}
The authors thank the referee for useful comments. This research has been supported by the funding for the ByoPiC project from the European Research Council (ERC) under the European Union's Horizon 2020 research and innovation programme grant agreement ERC-2015-AdG 695561. The authors acknowledge fruitful discussions with the members of the ByoPiC project (https://byopic.eu/team). 
This publication used observations obtained with {\bf \planck} (\url{http://www.esa.int/Planck}), an ESA science mission with instruments and contributions directly funded by ESA Member States, NASA, and Canada. It made use of the SZ-Cluster Database {\bf (http://szcluster-db.ias.u-psud.fr/)} operated by the Integrated Data and Operation Centre (IDOC) at the Institut d'Astrophysique Spatiale (IAS) under contract with CNES and CNRS. 

\end{acknowledgements}
\bibliographystyle{aa} 
\bibliography{dfil} 

\end{document}